\documentclass[sigconf, nonacm, authorversion]{acmart}
\AtBeginDocument{%
  \providecommand\BibTeX{{%
    \normalfont B\kern-0.5em{\scshape i\kern-0.25em b}\kern-0.8em\TeX}}}

\usepackage[T1]{fontenc}
\usepackage{multirow}
\usepackage{xspace}
\usepackage{hyperref}
\usepackage{graphicx}
\usepackage{url}
\usepackage[shortlabels]{enumitem}
\usepackage{listings}

\colorlet{punct}{red!60!black}
\definecolor{background}{HTML}{EEEEEE}
\definecolor{delim}{RGB}{20,105,176}
\colorlet{numb}{magenta!60!black}

\lstdefinelanguage{json}{
   numbers=left,
    numberstyle=\scriptsize,
   stepnumber=1,
   numbersep=8pt,
   numberstyle=\scriptsize,
   breaklines=true,
    literate=
     *{0}{{{\color{numb}0}}}{1}
      {1}{{{\color{numb}1}}}{1}
      {2}{{{\color{numb}2}}}{1}
      {3}{{{\color{numb}3}}}{1}
      {4}{{{\color{numb}4}}}{1}
      {5}{{{\color{numb}5}}}{1}
      {6}{{{\color{numb}6}}}{1}
      {7}{{{\color{numb}7}}}{1}
      {8}{{{\color{numb}8}}}{1}
      {9}{{{\color{numb}9}}}{1}
      {:}{{{\color{punct}{:}}}}{1}
      {,}{{{\color{punct}{,}}}}{1}
      {\{}{{{\color{delim}{\{}}}}{1}
      {\}}{{{\color{delim}{\}}}}}{1}
      {[}{{{\color{delim}{[}}}}{1}
      {]}{{{\color{delim}{]}}}}{1},
}

\lstdefinestyle{code}{
  language=json,
  float=tp,
  floatplacement=tbp,
  abovecaptionskip=-5pt,
  tabsize=2,
  numbersep=10pt
}

\definecolor{rhcol}{RGB}{66, 123, 245}
\definecolor{lightred}{HTML}{F4A582}
\newcommand{\mypara}[1]{\smallskip\noindent\textbf{#1}.}

\newcommand{\TOOL}[0]{\textsc{Z3Guide}\xspace} % [lisa 3/4] tool name change for anonymous review
\newcommand{\tool}[0]{\TOOL}
\newcommand{\riseforfun}[0]{\texttt{RiSE4Fun}\xspace}
\newcommand{\circledletter}[1]{\raisebox{-2pt}{\includegraphics[width=9.3pt]{attachments/#1.png}\Description{A circle with letter #1}}}

\newcommand{\undo}[0]{\raisebox{-2pt}{\includegraphics[width=9.3pt]{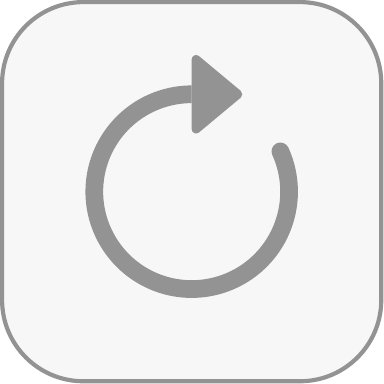}\Description{A button that undoes the reset action in the editor}}}

\newcommand{\reset}[0]{\raisebox{-2pt}{\includegraphics[width=9.3pt]{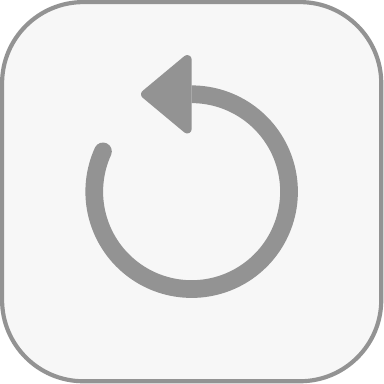}\Description{A button that resets changes in the editor}}}

\newcommand{\eg}{\emph{e.g.\@}\xspace}
\newcommand{\scode}[1]{{\texttt{\small #1}}}

\newcommand{\backtick}[0]{\textasciigrave{}}

\newcommand{\toollink}[0]{\url{https://github.com/microsoft/z3guide}}

\begin{document}

% \title{Human-Centered Web Experience for Education in Symbolic Logic Modeling}
%%% ICER title
% \title{A Framework for Teaching and Learning Logic Modeling via the Web}
%\title{Designing Tools for Teaching and Learning Logic Modeling} -- SIGCSE
% \title{\tool: Scalable Student-Centered Learning for Logic Modeling}
\title{\tool: A Scalable, Student-Centered, and Extensible Educational Environment for Logic Modeling}


\author{Ruanqianqian (Lisa) Huang}
\authornote{Work done while interning at Microsoft Research.}
\email{r6huang@ucsd.edu}
\orcid{1234-5678-9012}
\affiliation{
    \institution{UC San Diego}
    \city{La Jolla}
    \state{CA}
    \country{USA}
}

\author{Ayana Monroe}
\authornotemark[1]
\email{aam285@cornell.edu}
\affiliation{%
  \institution{Cornell University}
  \city{Ithaca}
  \state{NY}
  \country{USA}
}

\author{Peli de Halleux}
\email{jhalleux@microsoft.com}
\affiliation{%
 \institution{Microsoft Research}
 \city{Redmond}
 \state{WA}
 \country{USA}}

\author{Sorin Lerner}
\email{lerner@cs.ucsd.edu}
\affiliation{%
  \institution{UC San Diego}
  \city{La Jolla}
  \state{CA}
  \country{USA}
}

\author{Nikolaj Bj{\o}rner}
\email{nbjorner@microsoft.com}
\affiliation{%
  \institution{Microsoft Research}
  \city{Redmond}
 \state{WA}
 \country{USA}
}

\renewcommand{\shortauthors}{Huang, et al.}

\begin{abstract}

Constraint-satisfaction problems (CSPs) are ubiquitous, ranging from budgeting for grocery shopping to verifying software behavior. 
Logic modeling helps solve CSPs programmatically using SMT solvers. 
Despite its importance in many Computer Science disciplines, resources for teaching and learning logic modeling are scarce and scattered, and challenges remain in designing educational environments for logic modeling that are accessible and meet the needs of teachers and students. 
This paper explores how to design such an environment and probes the impact of the design on the learning experience.
From a need-finding interview study and a design iteration with teachers of logic modeling, we curated 10 design guidelines spanning three main requirements: providing easy access, supporting various educational modalities, and allowing extensions for customized pedagogical needs.
We implemented nine guidelines in \tool, an open-source browser-based tool. 
Using \tool in a logic modeling learning workshop with more than 100 students, we gathered positive feedback on its support for learning and identified opportunities for future improvements.

\end{abstract}




\maketitle

%% 
%% Insert sections here using the \input command.
%% For example, \input{intro} inserts the content in `intro.tex`.
%% Remember to create the file of which the content you are inserting.
\begin{sloppypar}
\section{Introduction}\label{sec:intro}

%\red{
%What is the lowest possible barrier for one to learn to use a symbolic calculator? How do we create a student-centered experience for this topic? 
%
%- low maintenance cost
%
%- cultivating community support
%
%- grounding in expert knowledge and student needs
%
%Points 1 and 2 can be solved technically. Point 3 needs to be solved via human input -- our contribution
%
%}

%\begin{enumerate}
%	\item Background: constraint solving problems, their applications, and logic modeling as a way to automate solving CSPs
%	\item State of the world: high barriers to systematically learn tools for solving CSPs; exacerbated due to covid-19 (increasing need for online resources)
%	\item Our take on this problem: crafting a student-centered learning environment via the web to reduce the barriers
%	\item Our implementation and what we found
%	\item Summary of contributions
%\end{enumerate}

\textbf{C}onstraint-\textbf{S}atisfaction \textbf{P}roblems (CSPs) involve finding an optimal solution under prescribed constraints.
Most decision making involves solving a CSP, from budgeting groceries to scheduling interviews.
Logic modeling is an approach to solving CSPs \emph{programmatically}, by translating constraints to a set of logical formulas and using ``\textbf{S}atisfiability \textbf{M}odulo \textbf{T}heories'' or SMT solvers to determine their satisfiability~\cite{freuder1997pursuit, gu1996algorithms}.
% Logic modeling,
% part of the Formal Methods discipline~\cite{woodcock2009formal},
% solves CSPs programmatically by translating constraints to a set of logical formulas and using ``satisfiability modulo theories'' or SMT solvers to determine their satisfiability~\cite{freuder1997pursuit, gu1996algorithms}.

With advances in SMT solvers like Z3~\cite{DBLP:conf/tacas/MouraB08} and their interfaces to widely-used programming languages such as C++ and Python, logic modeling is useful in many aspects of Computer Science, from Artificial Intelligence to Program Verification, and appears as a part of many standard computing curricula~\cite{curricula2013}. 
As such, there is a great need for accessible and student-centered educational environments for logic modeling, especially since the COVID-19 pandemic~\cite{runge2021teaching}.
However, creating an educational environment for logic modeling that is accessible, meets the needs of various teachers and students, and provides different learning modalities~\cite{folkestad2006formal}, imposes a design challenge~\cite{Davisetal} that prior work has yet to fulfill~\cite{pierceSeries, yurichev20xx, sage2024tutorial, alloy4fun, sagemathcell, ball2013increasing}.

Motivated by these concerns, we conducted a design exploration to identify design guidelines for an educational environment for logic modeling.
We interviewed six university faculty who have taught logic modeling with Z3~\cite{DBLP:conf/tacas/MouraB08}, which has been widely used for logic modeling, to understand user needs and technical requirements. We then developed a prototype learning environment, and iterated on the design with four faculty.
The design exploration surfaced 10 guidelines that suggest three main requirements for an educational environment for logic modeling: (1) providing easy access (2) supporting various educational modalities (3) allowing extensions for customized pedagogical needs.

We implemented nine of the 10 guidelines in \tool, a web-based tool for logic modeling.
\tool has an interactive textbook, a freeform editor, games, and access to external resources and the Z3 community, supporting both formal education and casual learning. 
\tool is 100\% client-side---with no server-side computation---and open-source\footnote{Maintained at \toollink}, minimizing its maintenance costs, allowing easy extension, customization, and contribution, and addressing the scalability and maintainability issues in its predecessor \riseforfun~\cite{ball2013increasing, rise4fun2021down}.
With \tool, students can learn the basics of logic modeling and programming with the Z3 API in various programming languages, by engaging in the following activities, all directly within a web browser: (1) interacting with textbook-like tutorials that include real-world problems and code examples (2) writing larger programs in a playground (3) solving logic puzzles.
Teachers can further use the tool for in-class demonstrations,  supplementary exercises, or custom extensions for their own pedagogical needs, such as promoting active learning~\cite{allen2005infusing}.

We used \tool in a three-hour online workshop where more than 100 participants learned logic modeling with \tool.
In a post-workshop survey ($N=21$), students rated \tool to be easy to use and found it enjoyable to learn logic modeling and Z3 using \tool.
We reflect on the design guidelines, in combination with our own usage experience, to inform future educational systems for logic modeling and beyond.

\section{Related Work}\label{sec:related-work}

\subsection{Background: Logic Modeling and SMTs}

%%% what is logic modeling
A Constraint-Satisfaction Problem (CSP) decides whether a given set of constraints can be satisfied.
Logic modeling solves a CSP programmatically by formally representing the CSP in a set of logical formulas and solving it using Satisfiability Modulo Theories.

%%% what are SMT solvers
Satisfiability Modulo Theories or SMT solvers support a formalism based on simply typed first-order logic with built-in theories for domains used in software. 
The theories include the theory of integer and real arithmetic, arithmetic over bit-vectors that correspond to the operations made by assembly code, as well as arrays and algebraic data types. 
These theories, along with the first-order logic, further enable modeling program behavior. 
As such, SMT solvers have been established since the early 2000s as a foundation for symbolic program analysis, verification, and testing.

%%% example of how logic modeling with SMT solvers works
We show an example of how logic modeling translates statements to first-order predicate formulas to be solved by the SMT solvers.
Given the following statements:
\begin{quote}
\centering
All humans are mortal. Socrates is a human. Therefore Socrates is mortal.
\end{quote}
The statements can be encoded as first-order predicate formulas:
\begin{align*}
\left(
(\forall x \ . \ \mathit{human}(x) \implies \mathit{mortal}(x)) \land
\mathit{human}(\mathit{Socrates})
\right)
\\
\implies
\mathit{mortal}(\mathit{Socrates})
\end{align*}
One could further translate the formulas to a format recognized by SMT solvers (e.g., SMTLIB~\cite{smtlib}) for their satisfaction to be checked.

%%% how logic modeling differs from imperative programming
Logic modeling with SMT solvers, grounded on first-order logic, is distinct from writing \emph{algorithmic} code in an imperative, functional, or object-oriented programming language. 
Unlike imperative programs, logical formulas have no notion of side effects.
Even when compared to functional programs, logical formulas cannot be evaluated.
Logic modeling is also distinct from modeling using linear algebra, which is widely used in operations research and data science disciplines.

%%% Z3 the SMT solver
Z3 is a state-of-the-art SMT solver widely used for logic modeling~\cite{DBLP:conf/tacas/MouraB08}.
Z3 supports many theories, reasoning with quantifiers, and customizing the solving process, enabling logic modeling in an extensive range of domains used in Formal Methods.
Our work resulted in \tool, an educational environment that assembles information about the full capability of Z3 starting from an elementary introduction to more advanced applications.

%%%%%
\subsection{Web Tools for Formal Methods Education}\label{subsec:formal-methods-in-edu}

Logic modeling is part of the Formal Methods discipline, which integrates mathematically rigorous techniques for the analysis and construction of systems~\cite{woodcock2009formal}.
There have been increased interests, especially since the COVID-19 pandemic~\cite{runge2021teaching}, in online curricula and tools that facilitate the education of Formal Methods.
Here we review web-based tools for the education of Formal Methods, including e-textbooks, programming environments, and tools that integrate them both.

%%% e-textbooks
E-textbooks allow for both guided and self-paced learning.
The Software Foundations series~\cite{pierceSeries} covers a broad range of Formal Methods topics from programming language theories to separation logic foundations.
SAT/SMT by Example~\cite{yurichev20xx} contains real-world examples of problems solvable by SAT/SMT tools.
Sage Tutorial~\cite{sage2024tutorial} introduces using Sage, a tool for modeling and solving math problems, with code examples.
These e-textbooks, however, are \emph{static}, requiring students to use one or more external environments to run the embedded code examples.
% : to use the embedded code examples, students need to go to one or more external environments (local or web-based) depending on how many programming languages/tools are involved.
%separately install related language tooling (e.g., the Coq proof assistant), potentially \emph{multiple} tools for multiple languages, and \emph{outside of the textbooks}.
%
% Unlike the static e-textbooks, 
Iltis~\cite{geck2018iltis,geck2022iltis} is \emph{interactive}, letting students learn and assess their learning of Logic and Theoretical Computer Science concepts via multiple-choice and short-response questions, but it does not have any code examples for applying the concepts to programming.
%
% However, Iltis does not provide any programming experience and is not tied to any programming tool, thus not supporting learning how to apply the concepts to programming.
%
Finally, for all these e-textbooks, an important limitation is that there are no effective ways for students to ask questions about any text sections or  examples.
%
%Our work results in \TOOL, a web environment for Z3~\cite{DBLP:conf/tacas/MouraB08} that includes an e-textbook with code examples while supporting programming \emph{within the textbook}, lifting the burden of programming environment setup. 

%%% programming environments
There are also web-based programming environments for the education of Formal Methods that contain no conceptual tutorial content.
%
%%%
KalkulierbaR~\cite{Kamburjan2021} provides a series of games that help practice concepts of logic modeling.
%
%%%
DiMO~\cite{dimo2021} supports using a SAT solver without installation.
Alloy4fun~\cite{macedo2021alloy4fun} enables writing programs of Alloy~\cite{jackson2019alloy}, a language for modeling software behavior, within the browser.
SageMathCell~\cite{sagemathcell} runs Sage code from the web, supplementing the Sage Tutorial~\cite{sage2024tutorial}.
One limitation of all of the tools above is that they rely on server backends, which induce costs of tool maintenance and scalability concerns.
A bigger limitation is their detachment from the tutorial materials of the corresponding languages/tools: while they ease writing programs in these languages/tools by not requiring software installation, the user could not learn about these languages/tools within the environments themselves.

% However, it relied on a server back-end that suffered from large traffic loads and adversarial attacks, which was part of the cause of the deprecation~\cite{githubrise4fun}.

%%% integrated environments
A web-based \emph{integrated environment} supports both learning and programming Formal Methods concepts by combining an e-textbook and a programming environment.
The most noticeable and closest to our work is \riseforfun~\cite{ball2013increasing}, a web environment for multiple logic modeling tools.
\riseforfun combined tutorials for logic modeling with a playground for each tool.
%for logic modeling with documentation for related concepts that many teachers previously used to facilitate teaching Formal Methods.
However, \riseforfun shared the same limitation as the e-textbooks~\cite{pierceSeries,yurichev20xx, sage2024tutorial, geck2018iltis} that there was no effective support for students to ask questions about the tutorial.
Also, like the environments above~\cite{Kamburjan2021, dimo2021, macedo2021alloy4fun, sagemathcell}, \riseforfun was limited in its implementation: it relied on a server backend that was vulnerable to attacks and large traffic~\cite{ball2013increasing}, and the maintenance costs were high due to its lack of extensibility~\cite{rise4fun2021down}.
For \riseforfun, these isues became severe to the point that it was discontinued~\cite{rise4fun2021down}.
Our system \tool addresses these limitations by adding shortcuts for asking tutorial-related questions in its interactive examples, implementing a completely client-side architecture, and enforcing extensibility.

More importantly, despite the variations of web-based educational tools for Formal Methods, comparatively few works explore how to design them to effectively support teaching and learning.
\citet{runge2021teaching} found through two qualitative studies with students of Formal Methods that, for these tools to facilitate online teaching, they must support various educational modalities and be interactive, accessible, and engaging.
Our work reveals more pedagogical needs for an educational tool for logic modeling through a design exploration and a workshop evaluation.
%

%
%From two studies with students, Runge et al.~\cite{runge2021teaching} found that web-based tools for Formal Methods must be interactive and accessible -- in both configuration and affordability -- for them to be easily incorporated into teaching.
%We precisely designed such an experience in \TOOL based on the needs of teachers.

%%
%The tools above are either separate environments from a textbook~\cite{Kamburjan2021, dimo2021, macedo2021alloy4fun}, or unable to handle large traffic or provide a stable service due to implementation limitations~\cite{rise4fun}.
%%
%Our prototype \TOOL distinguishes itself by enabling programming directly within a textbook and providing a stable, scalable service via a client-only infrastructure.
%%
%In addition, we developed \TOOL with a human-centered design, the first to our knowledge in technologies for Formal Methods education.

%

\subsection{Web Tools for Programming Education}\label{subsec:web-based-programming-tools}

Beyond logic modeling and Formal Methods, there is a rich line of web-based tools that support programming education.
Online interactive textbooks like Stepik~\cite{stepik2023} and Runestone~\cite{ericson2020runestone} let students learn programming concepts and complete programming practices in the browser.
There are also various programming environments covering a broad range of topics.
Python Tutor~\cite{guo2013online} 
% is one of the most widely cited environments that 
has allowed millions of users to program, share, and visualize Python code in the browser, and led to design guidelines~\cite{guoTenMillionUsers2021} for scalable research software in academia.
Inspired by Python Tutor, Pandas Tutor and Tidy Data Tutor~\cite{lau2023datascience} support data science education at scale.
Godbolt~\cite{godboltCompilerExplorer} allows students learning compilers to compare compilation outputs from different compilers without any installation, which is usually platform- and architecture-dependent.
CS50 uses GitHub Codespaces~\cite{codespaces} to support introductory programming without local configuration~\cite{cs50dev}.
Finally, integrated environments like Ed~\cite{ed2023} combine digital textbooks, programming experience, and online discussions.
Our work is complementary to these efforts by focusing on the education of logic modeling and adopting a human-centered design.

\section{Design Exploration with Teachers}\label{sec:design-exploration}

% What are the user needs and challenges in using software tools for the education of logic modeling? 
%
% To understand the design space of an educational tool for logic modeling, 

To derive design guidelines for an educational logic modeling tool, we conducted a two-stage exploration: a need-finding interview study with six university faculty (\autoref{subsec:need-finding-interviews}), and a design iteration with four previously interviewed faculty on a prototype we built based on the interview findings
% , to further reveal content and system requirements and refine the design goals
(\autoref{subsec:cognitive-walkthroughs}).
Both studies were run by the first two authors via web conferencing, involved no compensation, and received IRB approval.
This section reports the methodology of each study.
Detailed findings are reported in~\autoref{sec:design-goals-of-tool}.

%\todo{
%relate to the following questions:
%``How does one build an educational environment for logic modeling that meets the needs of the teachers and the students? When designing such a system, how does one balance user experience, maintenance effort, and scalability?''
%}

% To explore the design space for an educational tool for logic modeling, 
% we interviewed six university faculty.
% %
% Based on the interview findings, we implemented a prototype system.
% %
% We iterated the prototype design with four of the previously interviewed faculty to further reveal usability requirements and refine the design goals.
% %
% We summarized the findings from the entire design exploration---interviews and design iteration---into three design goals.
% %
% Both studies in the design exploration were run by two authors via web conferencing, involved no compensation, and received IRB approval.
% %
% We describe the methodology of each study in~\autoref{subsec:need-finding-interviews} and~\autoref{subsec:cognitive-walkthroughs}, and all results in~\autoref{subsec:design-goals-of-tool}.

\begin{table}[t]
    \centering
    \footnotesize
    \caption{Need-finding study participants. ``Teaching Focus'' is the main topic of their classes where logic modeling is~taught. }
    \begin{tabular*}{\linewidth}{@{\extracolsep{\fill}}lllll@{}}
        \toprule
        ID & Gender & Title                  & Main Area of Expertise & Teaching Focus \\
        \midrule
        P1 & M   &  Assoc. Prof.   &   Compilers  & Compilers            \\
        P2 & F    & Assoc. Prof.          & Formal Methods    &   Verification           \\
        P3 & M       & Prof.   &  Verification    &  Verification           \\
        P4 & M & Prof.   &  Automated Reasoning   &   Verification     \\
        P5 & F &  Assoc. Prof.   &  Program Synthesis   & SAT/SMT            \\
        P6 & M       & Asst. Prof. &   Formal Methods                 &    Verification         \\
        \bottomrule
        \end{tabular*}\label{tab:need-finding-participants}
    \Description{
    Table with seven rows, the first row as the header row, and five columns, the first column as the header column. The table is as follows from left to right:
    ID Gender Title Main Area of Expertise Teaching Focus
    P1 M Assoc. Prof. Compilers Compilers
    P2 F Assoc. Prof. Formal Methods Verification
    P3 M Prof. Verification Verification
    P4 M Prof. Automated Reasoning Verification
    P5 F Assoc. Prof. Program Synthesis SAT/SMT
    P6 M Asst. Prof. Formal Methods Verification
    }
\end{table}

% \begin{table*}[t]
%     \centering
%     \small
%     \caption{Need-finding interview participants. ``Teaching Focus'' refers to the main topics of their classes where logic modeling is taught. ``Design Iteration?'' indicates whether a participant also participated in our design iteration study.}
%     \begin{tabular}{llllll}
%         \toprule
%         ID & Gender & Title                  & Main Area of Expertise & Teaching Focus & Design Iteration? \\
%         \midrule
%         P1 & M   &  Assoc. Prof.   &   Compilers  & Compilers  & Y          \\
%         P2 & F    & Assoc. Prof.          & Formal Methods    &   Verification  & Y         \\
%         P3 & M       & Prof.   &  Verification    &  Verification  & -         \\
%         P4 & M & Prof.   &  Automated Reasoning   &   Verification & Y     \\
%         P5 & F &  Assoc. Prof.   &  Program Synthesis   & SAT/SMT & Y           \\
%         P6 & M       & Asst. Prof. &   Formal Methods                 &    Verification & -        \\
%         \bottomrule
%         \end{tabular}\label{tab:need-finding-participants}
% \end{table*}

\subsection{Need-Finding Interview Study}\label{subsec:need-finding-interviews}
%
% \mypara{Participants}
%
We recruited participants by emailing faculty with publications and teaching experience in logic modeling according to their websites.
We required that participants have experience with \riseforfun~\cite{ball2013increasing}, a deprecated web environment for logic modeling that had been widely used by teachers.
Eventually, six faculty from six institutions across five countries 
% who had taught logic modeling with Z3 and \riseforfun 
gave their consent to participate.
% Six faculty from six institutions across five countries gave their consent to participate.
%
% The participants are from six institutions across five countries.
%
\autoref{tab:need-finding-participants} shows their demographics.
%
% All participants had taught logic modeling with Z3 and \riseforfun~\cite{rise4fun}, a popular yet deprecated web environment for the topic.
%
We deem the sample size of six participants reasonable due to the small size of the logic modeling community.
% in comparison to the larger Computer Science~community.

%%%% Procedure
% \mypara{Procedure}
%
Each participant went through a 60-minute semi-structured interview.
%anywhere from 40 to 70 minutes.
% via web conferencing. 
%
We first asked how they taught logic modeling and related tools.
% had integrated logic modeling and related tools into teaching.%the classes they had taught. 
%
Then, we asked questions from four categories:
(1) the integration of \riseforfun into the curriculum
(2) failures and successes of the integration in supporting their and/or the students' needs
% (2) how the integration was successful
% (3) how \riseforfun had failed their and/or the students' needs
(3) challenges faced teaching logic modeling and Z3
(4) expectations for a better tool.
%
% % Each interview contained questions from five categories: 
% (1) How and why \riseforfun had been 
% % was another tool reminiscent of the \TOOL called the rise4fun guide \cite{rise4fun},  
% integrated into the curriculum and in-class teaching activities;
% (2) How \riseforfun had been successful in such integration;
% (3) How \riseforfun had failed their and/or the students' needs;
% (4) What issues had arisen when teaching logic modeling and Z3;
% (5) How a tool with similar goals could be better.
%
% After this first question, each participant was able to speak for as long as they would like on the 5 main topics referenced previously. 
Each category had a list of questions that interviewers could ask out of order and follow up on when necessary.
%
% \red{A full list of the questions could be found in LINK }\rhnote{do we want to attach an anonymized document as appendices don't seem to be allowed?}.
%
% Two authors ran all the interviews.
%
% Participants received no compensation, and our protocol received IRB approval.

% \mypara{Data}
%
We recorded the audio and video of each interview.
The web-conferencing software  transcribed the audio, and the first two authors fixed transcript errors by revisiting the video recordings.
%
% Given the exploratory nature of the study, 
We analyzed the data using thematic analysis~\cite{braunUsingThematicAnalysis2006}.
The second author coded all data, and
the first author coded 25\% of the data to compute percentage agreement~\cite{syedGuidelinesEstablishingReliability2015} and establish reliability.
% computed percentage agreement~\cite{syedGuidelinesEstablishingReliability2015} using 25\% of each  coded transcript to establish reliability of the analysis.
%
%For example, if a transcript has 100 coded data, then the coder randomly samples 25 coded data along with the set of all available codes, removes the codes from the data, and asks the other coder to assign codes.
%%
%We consider that the two coders agreed upon a data if they had at least one code in common. 
%
Eventually, the coders agreed upon 88\% of the sampled data.
% from the interview study.
%\rh{consider recomputing for Cohen's kappa instead -- this might be a stronger measure}
%
The first author then grouped the codes into themes, which are reported as paragraph headings in \autoref{sec:design-goals-of-tool}.

%\mypara{Summary of Findings}
%
%
%Raw findings
%\begin{enumerate}
%    \item Z3 and rise4fun are not very fast (P1, P2)
%    \item rise4fun lacks instructions for applying Z3 to their work (P1)
%    \item no feedback on execution status in rise4fun; sometimes it is reasonable that verification tools run for a long time, but people need feedback (P1)
%    \item it would be great to make each step transparent just like in Godbolt (P1); better visual-ish explanation for model behavior (P2)  
%    \item content: students need help for quantifiers, optimization (P1)
%    \item for learning environment, web-based is the way to go; really like the installation and configuration free experience in rise4fun (P1, P2)
%    \item logic background content is important (P1, P3)
%    \item easy, asynchronous way of asking for help (P1, P3)
%    \item teachers don't care about security in maintenance (P1)
%    \item gamification and realistic examples are especially important (P1, P3) because ``verification is a very novel concept'' (P2)
%    \item students used rise4fun as a debugging environment (like a REPL), for them to test out smaller pieces of a potentially larger code (P2)
%    \item integration with existing tools their institutions use (e.g., Google Suite)
%\end{enumerate}

\subsection{Design Iteration}\label{subsec:cognitive-walkthroughs}

%
% % \mypara{Participants}
%
Based on findings from the interviews, we developed a prototype similar to our final design (\autoref{sec:design}).
We re-recruited the interview participants to evaluate the initial prototype per the iterative nature of the study.
Four participants (P1, P2, P4, P5 in \autoref{tab:need-finding-participants}) agreed to be in the prototype evaluation. 
%

% \mypara{Procedure}
%
Each 45-minute evaluation consisted of a 15-minute open-ended exploration 
% to obtain the participant's opinion on the prototype 
and a 30-minute cognitive walkthrough~\cite{lewis1997cognitive}, a commonly used technique for evaluating early prototypes.
% to assess its usability.
%
% Cognitive walkthroughs are commonly used for evaluating early design prototypes: 
Anyone experienced in UX research or the corresponding domains could perform cognitive walkthroughs while revealing usability problems via realistic use cases.
In our study, the participants are experts in the education of logic modeling: they are familiar with the topic and the kinds of problems the students could encounter.
%
% Each participant first spent 15 minutes exploring the features of the prototype, asking the interviewers clarifying questions, commenting on things they noticed, and summarizing how they understood the prototype.

During the initial exploration, each participant commented on things they noticed, asked clarifying questions, and summarized how they understood the prototype.
In the cognitive walkthrough, they picked one topic in the prototype they would cover (e.g., ``Logic'') when teaching a related course (\eg, software verification) and used the tool following an imaginary scenario:
% They then picked one section of the documentation (e.g., ``Logic'') and spent 30 minutes interacting with the interface following an imaginary scenario:
\begingroup
    \addtolength\leftmargini{-0.15in}
    \begin{quote}
    ``You were a student in [a specified course] learning logic modeling and Z3, and you were asked to use the tool as supplementary reading and exercises. How would you use it to learn concepts in [the selected section]?''
    \end{quote}
\endgroup
%
% \begin{quote}
%     `If you were a student in your software verification class learning logic modeling and Z3, and you were asked to use the tool as a supplementary textbook for the course to read the included material and write programs with Z3, how would you use the tool to learn concepts in the ``Logic'' section?
%     % ``You would like your student to learn logic modeling and Z3 for your software verification class. You have prepared lecture material and homework assignments for related topics, but you would like to use this tool as the ``supplementary textbook'' for these topics. You expect your students to read material and write programs within this tool without installing Z3 locally. Now, pretend that you were one of your students, and go to the ``Logic'' section. How would you use this interface to learn logic with Z3?''
% \end{quote} 
%
\noindent
For each action the participants performed, we asked how they expected the interface to react, whether the existing design matched their expectations, and their ideal design and why.
%
% All sessions were conducted by two authors.
%\rhnote{Clarify who the authors are in camera-ready}

%\red{
%%
%Note that, in our study, the participants who performed the walkthroughs are \emph{teachers} rather than students of logic modeling.
%%
%The decision was made partially because we had little access to students of logic modeling during the time of the study, \ie, summer breaks.
%%
%However, we deem our approach still valid because of the following: (1) cognitive walkthroughs can be done by anyone (2) these participants have taught logic modeling multiple times and know what a typical student could behave.
%%
%}

% \mypara{Data Collection and Analysis}
%
We recorded the audio, video, and the participant's screen of each session, and noted down interactions with the prototype.
We used the same audio transcription approach in~\autoref{subsec:need-finding-interviews}.
% We transcribed the audio recordings following the same approach in~\autoref{subsec:need-finding-interviews}.
%
Relating to the findings from~\autoref{subsec:need-finding-interviews}, the first author used top-down coding to code the transcripts and notes of interactions.
%
% We computed the inter-coder agreement using the same approach in~\autoref{subsec:need-finding-interviews} and received 87.5\% agreement on the sampled data.
Like in~\autoref{subsec:need-finding-interviews}, the first two authors used 25\% of the coded data to compute agreement, which was 87.5\%.
The authors agreed upon themes emerging from the codes reported as paragraph headings in \autoref{sec:design-goals-of-tool}.

\section{Initial Design Guidelines}\label{sec:design-goals-of-tool}

%
% Based on findings from the interviews (\autoref{subsec:need-finding-interviews}) and the design iteration (\autoref{subsec:cognitive-walkthroughs}),
We derived 10 design guidelines (denoted as \textbf{D}s below) under three categories for educational tools for logic modeling from the interviews and design iterations (\autoref{sec:design-exploration}).
% that drive our tool design (\autoref{sec:design}). 
%
% \textbf{D1} is further subdivided into \textbf{D1a}, \textbf{D1b}, etc., similarly for \textbf{D2} and \textbf{D3}.
%

%%%
\mypara{Providing Easy Access}
All participants considered easy access to the tool---regardless of educational setting and role---a priority, particularly with regard to programming experience.

%\rh{BS: D1.1, ... for the sub-bullets for easier references in Section 4?}
%
% NSB: use instead mypara with $\bullet$ to indicate sub-enumeration?
% I have applied the change
% to this paragraph but not
% the next ones. Lisa has to
% make the call what to do
% (apply elsewhere or revert).

%\begin{enumerate}[label=, leftmargin=2pt]
    
 %       \item
 \mypara{\textbf{D1: \textit{Fast Execution}}} 
        {
         In the interview, five participants recalled unpredictable slowdowns in \riseforfun that interrupted lectures. 
         They remarked that \textit{interactive speed} is necessary for the tool to facilitate effective classroom interactions.
%         the educational flow for students and teachers.
         
         During the design iteration, all participants liked the \emph{``snappy''} (P4) speed of our prototype most of the time. However, the tool froze for P2 and P5 at the end of their sessions after several code executions due to a memory leak bug that we later fixed.
%    % %
%    % We resolved both bugs before deploying it for the workshop evaluation (\autoref{subsec:workshop-evaluation}).
        %
        % While the interruptions seemed random for the participants, 
        % % they were actually caused by over-contention for the shared servers when too many users would get onto the system
        % they were actually caused by the large number of requests for the shared \riseforfun servers when too many users ran their code 
        % -- a situation that our framework should avoid. 
        % With \TOOL, this problem is addressed by running the system on the client's side, inside the browser. --> we don't mention our system until later in the section
        }

%                \item
\mypara{\textbf{D2: \textit{Code Sharing}}} 
        {
        Four interview participants (P1, P2, P3, P6) deemed it important to easily share code snippets with others.
        With \riseforfun, one could share permalinks with others to recreate the state of the tool and their code.
        % \riseforfun had ``permalinks'' that could be shared and opened by others to recreate the state of the code and tool. 
        %
        While this feature might not be essential (due to workarounds like copying/pasting), the four participants considered it to be useful, as \emph{``people send links to each other all the time [to] teach each other'}' (P1).

        Our prototype did not enable direct code sharing or a shortcut for copying code snippets.
        However, P2 attempted to copy code snippets across editors multiple times when using the prototype with the select-all and copy keyboard shortcuts, and P1 suggested that students should be able to share/replicate both a code snippet and its execution state with their classmates or teachers.
        %
        % [logic modeling concepts]'' (P1).
        % \rhnote{some quote (segments) would be helpful here}
        % \nsbnote{the quote below appears to compile several quotes. One snippet could be sufficient, "People send links to each other all the time and they teach each other...". Drop the "you know"}
        % \nsbnote{the item bullet does not show up here. I would just not use it for any of the quotes}
        % \begin{quote}
        %       That’s mandatory and the users just want this feature like... People send links to each other all the time and they teach each other, you know? Ohh here's the link. And then someone edits it a little bit and something another one ... they trust... Somehow these websites are, are safe. - P1
        % \end{quote}
              
 }
        
 %       \item
 \mypara{\textbf{D3: \textit{Editing Support}}}
        {
%         A minority of participants commented that RiSE4Fun lacked visual
% indicators in the text editor and output display, like syntax highlighting or visually formatting long out. To them, adding these features in a new tool would elevate the experience of teachers and learners, and reduce the number of errors that they make.
        % A minority of \rhnote{say "Two" if there were only two} participants (P4, P6) \ayana{hotel wifi slow, Arie should be p3 -- [lisa: p4 actually] check at ucsd)} 
        In the need-finding interview, P4 and P6 found it necessary to have visual indicators in the code editor and output display, such as syntax highlighting and visually formatting long output, which \riseforfun lacked.
        % visual indicators in the code editor and output display, like syntax highlighting or visually formatting long output. 
        They believed that these features would help students catch errors earlier on and understand the output.
        % , and reduce the amount of errors they make.
        %
        Compared to debugging algorithmic code, debugging logic modeling code has been a hard problem~\cite{goualard1999visualization, leo2017debugging}, but part of it could be mitigated with appropriate editor support for getting the syntax correct and avoid \emph{``the [output] given to the students [not being] straightforward''} (P6).
%        In particular, if the code syntax ``is incorrect'' in the first place and such information is not communicated through the editor, then ``the [output] given to the students [might not be] straightforward'' (P6).
        %

        The design iteration further reinforced the necessity of appropriate editing support.
        Our prototype lacked syntax highlighting and autocompletion, while Z3 could be written in SMTLIB, a format that adopts a Lisp-like syntax with a heavy use of parentheses.
        As such, every participant made at least one mistake in balancing parentheses that took a long time to debug during the cognitive walkthrough due to the lack of IDE support.
        In addition, P4 further found the localization of error messages difficult because the editor did not have line numbers.
        Finally, all four participants demanded that the tool support resetting an edited snippet to its original content and reverting any accidental reset.
        }

%    \end{enumerate}

\mypara{Supporting Multi-Modal Education}
%
% Participants deemed it important for the tool to engage and support learners from four aspects.
Participants proposed engaging and supporting students via the tool from five aspects.
%
% Our studies surfaced four kinds of learning support.

%    \begin{enumerate}[label=, leftmargin=4pt]

            %\item 
\mypara{\textbf{D4: \textit{Small Examples}}}
            {
            % The prioritization of small code snippets and changes to code was requested by a majority of participants.
            Most interview participants
            % in the need-finding interviews
            requested that the tool explain logic modeling concepts through small code examples.
            % use small code examples for explaining
            %
            % By building infrastructure that prioritizes this programming technique and code snippets of this size, 
            Participants believed that by building content with small code examples,
           the tool would encourage students to run the examples and lead to a better understanding of the concepts.
            % that highlights specific ideas integral to understanding formal methods of programming. 
            %
            In addition, the small examples might also encourage students to make edits, \emph{``play around, and [re-execute]''} (P3), which could deepen their understanding of the connection between the code and the output.

            The design iteration confirmed the benefits of executable and editable code examples, as all participants tweaked and executed the examples that came with the prototype while going through the material.
            %
            % Our prototype provided tutorial materials embedded with editable and executable code examples, and all participants tweaked and executed the examples going through the materials.
            %
            P2 further pointed out that allowing the user to run each code snippet on the fly could avoid \emph{``the hassle of copying and pasting code examples''} into an external editor.
            She imagined letting her students \emph{``run [each code example] themselves''} in class to reinforce their understanding of the related topics.
            Finally, participants suggested having code examples with intentional errors so students could learn about repairing them.
            }

%            \item
            \mypara{\textbf{D5: \textit{Freeform and Exploratory Programming}}}
            Four interview participants (P2, P4, P5, P6) wanted \emph{``a simple playground where [students] can try certain things out''} (P4) in the tool.
            	P6 deemed a freeform editor also important for the teacher to quickly demo some logic modeling code (i.e., \emph{live coding}~\cite{selvarajLiveCodingReview2021}) for the students to follow along and further explore.
            	%
			% To do so with Z3, he would need to make sure both he and his students had some Z3 package installed, which would be a hassle at the beginning of the course.

            	%
            	Our prototype came with a freeform editor that, compared to other concept-related interactive examples, was on its own page and not attached to any concept.
            In her walkthrough,	P5 demonstrated how a student could use the editor to create logic formulas and solve an example problem in her course slides.
            % \footnote{We detailed part of their demonstration in one of the usage scenarios in \autoref{subsec:usage-scenarios}.}
            	However, the freeform editor was not editable until after the user ran the sample code inside, which the participants found confusing and contrary to its intention for freeform explorations.
            This design inherited that of the concept-related examples.
        %    %
                While all participants liked this design for those examples,
                they suggested that the freeform editor be always editable to encourage exploratory programming.
                The free editor in the prototype was also small, which P1 thought should be larger to encourage various sizes of exploration.

\mypara{\textbf{D6: \textit{Gamification}}}
            {
            Two interview participants (P1, P3) believed that games would benefit  logic modeling students. 
            P1 particularly referred to competitions (a form of games) for compiler optimization: he would use competitions in his class to encourage students to write compilation code as optimized as possible, then release the competitions to the public with rewards for the winner, and people in the compilers community \emph{``love these kinds of competitions.''}
            % As a comparator, participants highlighted competitions organized by their respective practicing communities. 
            %For instance, P1 brought up a competition hosted in the compiler community. 
            He suggested that such competitions could be part of the tool to help practice concepts and engage students.

            Our prototype did not include any games, and P1 reiterated the importance of having alternative forms of educational materials beyond readings and exercises in the tool.
        }

%            \item
\mypara{\textbf{D7: \textit{References to Basic Concepts in Logic}}}
            {  
            Logic modeling has a basic logic prerequisite, but according to the interview participants, students still occasionally need references to these concepts.
            % Several were noted as confusing for students, however the only recurring concept was basic logic. 
            
            Our prototype lacked such references, to which P4 remarked that most students at the start of the academic term \emph{``[didn't] remember anything~[\ldots]~about basic logic''} although \emph{``everybody had taken [the prerequisite course]''} (P4).
            % In the words of P4 when they get to their class and they ask them what they remember about basic logic "Most of them don't remember anything, so like I asked them, you know, do you know any of those concepts? They say no. 
            % And it's like, who's taken 250? Everybody took 250." 
            % 
            This quote exemplifies the need new logic modeling students often have for a basic logic concepts refresher.
            % As exemplified by this quote, when many students get to classes involving SMTLIB or Logic Modeling, enough time has passed between that class and their first exposure to logic for students to forget. 
            % 
            % Often, a review of simple logic concepts is necessary since they have been forgotten. 
            % 
            It is thus beneficial to provide basic logic references in the tool.
            % that are quite often forgotten by learners.
            % \item Games and Visuals: Another suggestion commonly proposed by participants was the addition of games and visuals that can serve as entertainment and learning devices for students. In current learning processes, participants referenced games that they currently Adding games that students can recreate and then play will inspire their learning, as it is  
            }

%            \item
\mypara{\textbf{D8: \textit{Question Asking}}}
            {
            Four interview participants (P1, P3, P4, P5) mentioned that students learn through asking questions and obtaining answers~\cite{allen2005infusing}, but many of them feel uncomfortable asking or answering questions during class.
            Asynchronous question-asking thus became a feature of interest.
            % Participants were thus interested in supporting asynchronous question asking through our framework.
            % In order to support students who are less likely to speak up in class for any reason, participants expressed an interest in, providing a mechanism for students to ask questions asynchronously. 
            %
            For example, P1 used Slack~\cite{Slack} in his class so that students could ask questions and receive help from the teaching staff and other students asynchronously.
            % A tactic that works well for one participant are Slack groups that allow students to ask questions to them or teaching assistants.% 
            %
            The participants would particularly like the tool to support asking questions about individual code snippets because most questions involve specific code examples.
            Moreover, they saw the potential of native question-asking inside the tool for casual learners to receive help from the entire logic modeling community.
            % Moreover, they acknowledged the potential benefits of framework-supported question asking to learners outside the traditional classroom setting because these learners could receive help from the entire logic modeling community.
            % Many participants asked that the tool allows students to ask questions about particular code snippets since this may get less attention in the classroom. This online query format will also benefit learners outside the classroom as well.

            	In our prototype, each code example came with a ``Discuss'' button that took the user to the GitHub Discussion of the Z3 repository, which is active with people asking/answering questions about Z3 and logic modeling.
            	The button, which was next to the ``Run'' button for each code snippet, received compliments for its functionality (P2, P5) and complaints about its appearance (P1).
            	P1 suggested a redesign of the button for it to be noticeable but not distracting.
            }
%        \end{enumerate}

\mypara{Allowing for Extensions}
Participants hoped for open source in both the tool and related educational resources.
% Participants would also like to see extensibility in the tool, in both resources for education in logic modeling and the tool itself.

% \begin{enumerate}[label=, leftmargin=4pt]
            
    %\item 
\mypara{\textbf{D9: \textit{Resources Sharing}}} 
    {
    Four interview participants (P1, P2, P3, P5) said that teachers of logic modeling typically share teaching resources (e.g., course slides) with one another and reuse existing ones.
    % build new ones on top of existing ones.
    % According to participants, the norm for teachers in the Logic Modeling education is to share resources and modify them for their needs. 
    %
    However, there was no dedicated space for the sharing of such resources.
    Participants saw an opportunity in an educational tool for logic modeling for \emph{``sharing and publishing [these teaching resources] systematically''} (P2) that makes them accessible, credits their authors, and enables community-driven improvements.
    % by making these resources accessible and allowing others to post their additions and changes while giving credit to the original. 
    %
    % Other participants proposed similar ideas and agreed with the potential benefits of this resource sharing approach.
    % \rhnote{I commented out the legality aspect as I'm not sure how it connects with what we proposed in the end}
    % While some other participants proposed similar ideas and agreed with the potential benefits, there were concerns about the legality of this resource sharing approach.
    % % While a majority of participants felt some version of this mechanism would be beneficial, there were concerns about the legality of this practice in general. 
    % It is thus necessary for our framework to support sharing teaching resources while respecting the rights of their  creators.
    % The development of a system that supports this common practice and respects the rights of educational content creators is necessary for \TOOL. 
    % Besides teaching resources, community-built tools in the Z3 ecosystem could be also shared and advertised in the tool.
    The tool could also serve as a platform for informing users of related technologies.
    % tools available in the Z3 ecosystem.
    %
    For example, user propagators are techniques that allow users to write custom theory extensions for the Z3 solver~\cite{bjorner2023satisfiability}.
    %
    % User propagators are an example of such tools, which allow users to write custom theory extensions for Z3 \cite{bjorner2023satisfiability}.
    % For example, one important part of Z3 is User Propagators, which allow the implementation of custom theory solvers \cite{bjorner2023satisfiability}. 
    %
    P3 and P4 had created user propagators, and they would like the work built by themselves and others to be available to teachers and students. 
    In particular, P3 suggested that there could be a centralized space for Z3 users to access these propagators, whether they are work-in-progress or ready-to-use.
    
	Our prototype provided links to other Z3- and logic modeling-related resources in the bottom, which caught the attention of P1 and P2, both of whom found them to be useful.
    }

    %\item 
\mypara{\textbf{D10: \textit{Tool Extensibility}}} 
    {
    Three interview participants (P2, P3, P4) desired the ability for teachers to easily extend the tool for their pedagogical needs.
    % , such as adding new content.
    % add new content that is in line with their lectures instead of pre-generated \TOOL content. 
    %
    An extensible tool could allow teachers to use its technical infrastructure while adding their own content, including text and code examples.
    % The tool needs to support the teachers that wish to make use of \TOOL's technical infrastructure and supplement their own examples which should not be difficult since the tool's content could be modified with GitHub. 
    %
    % Another part of extensibility that the participants hoped for is allowing
    Participants also hoped for integrations with existing learning platforms for activities such as grading and assignment generation.
    Ideally, the tool should enable connections to these services, e.g., at the source code level.
    % Another part of extensibility is allowing integrations with existing learning platforms that support functions like grading or assignment generation. Some sort of connection needs to be in place in case teachers chose to link \TOOL and their existing educational platforms.
    
    Although our prototype did not implement extensions facing the teachers, in the design iteration we mentioned that the tool would be open source on GitHub for contributions and extensions, about which all participants were excited.
    }

%\end{enumerate}

% \input{03_design_exploration}
% \input{04_design_goals}
\section{Design of \tool}\label{sec:design}

We designed \tool, a web-based tool for the education of logic modeling with the Z3 SMT solver~\cite{DBLP:conf/tacas/MouraB08} that is scalable, student-centered, and extensible, addressing all design guidelines (\autoref{sec:design-goals-of-tool}) except \textbf{D2}.
\autoref{subsec:usage-scenarios} previews the design via usage scenarios.
\autoref{subsec:access-and-programming}--\autoref{subsec:customization-and-extension} each details how we accomplished each of the three categories of design guidelines reported in \autoref{sec:design-goals-of-tool}.

%We refer to components in~\autoref{fig:overview} to describe the design and implementation of \TOOL based on the design~goals.}
% based on the three design goals (\textbf{D1}, \textbf{D2}, \textbf{D3}) derived from our design exploration in~\autoref{sec:design-exploration}.

\begin{figure*}[th]
	\centering
	\includegraphics[width=\textwidth]{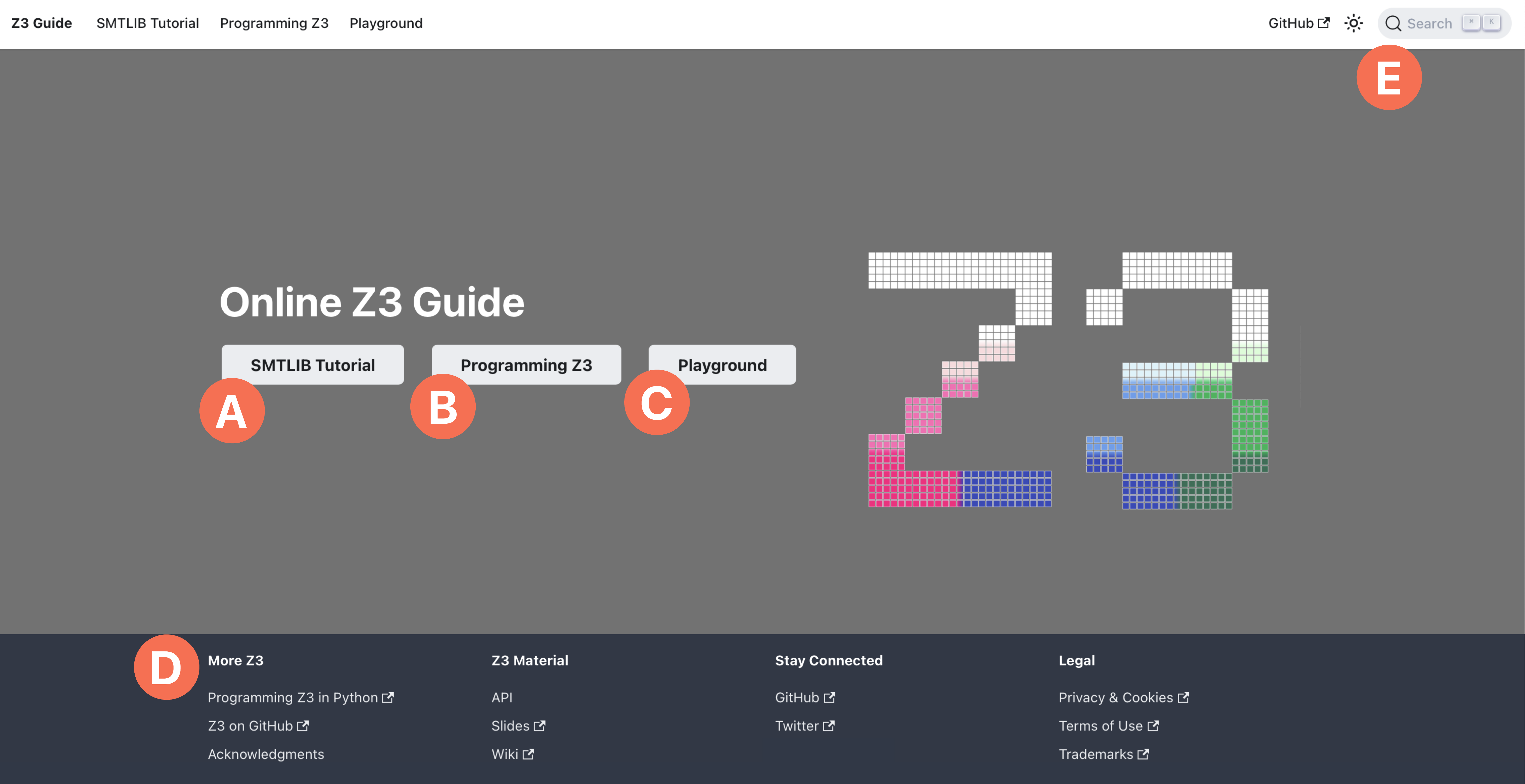}
    \Description{
    Screenshot of the homepage of Z3Guide. There are three visual rows of the content. The top row has two sections. The left section contains four hyperlinks that go to the homepage (the page shown), SMTLIB Tutorial, Programming Z3, and Playground, respectively. The right section contains three components: a hyperlink that goes to the github repository of Z3Guide, a button that toggles between the light and dark theme, and a search bar that enables searching content within Z3Guide. There is an E label attached to the search bar. The middle row of the homepage contains three buttons and a logo of the tool. There is an A label on the first button that, upon clicking, goes to SMTLIB Tutorial. There is a B label on the second button that, upon clicking, goes to Programming Z3. There is a C label on the third button that, upon clicking, goes to Playground. The bottom row of the homepage contains several hyperlinks to external resources related to the tool, with a D label attached to the row.
    }
    \caption{
    % Learning and teaching logic modeling and Z3 with \TOOL, which contains: tutorials in different language syntaxes, code editors with pre-filled examples, games, and connections to the online community and external resources. 
    The \TOOL interface:
    \emph{SMTLIB Tutorial} (\circledletter{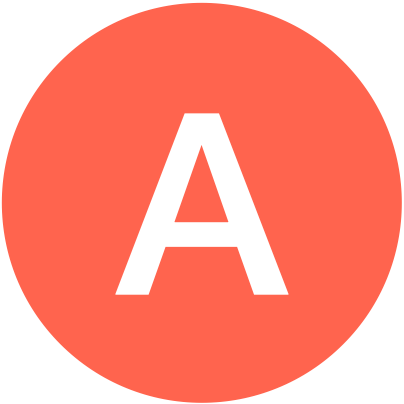}), upon pressing, shows \autoref{fig:smtlib} that includes tutorial content and editable code examples for logic modeling in SMTLIB format.
    \emph{Programming Z3} (\circledletter{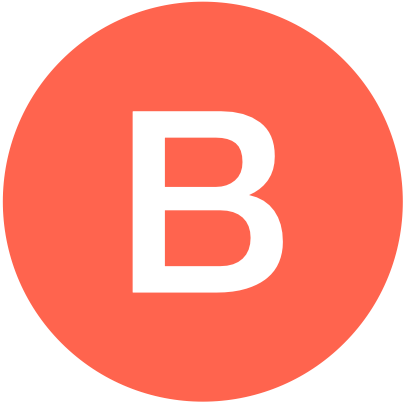}), upon pressing, shows logic modeling with Z3 bindings in JavaScript (\autoref{fig:js}) and Python.
    \emph{Playground} (\circledletter{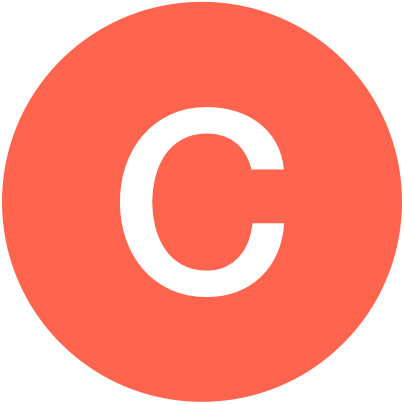}), upon pressing, shows logic formula guessing games (\autoref{fig:secret-formula}) and a freeform editor (right half of \autoref{fig:freeform-editor}).
    It also comes with links to external resources (\circledletter{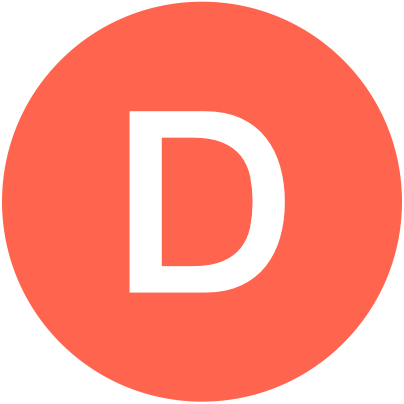}) and a built-in search (\circledletter{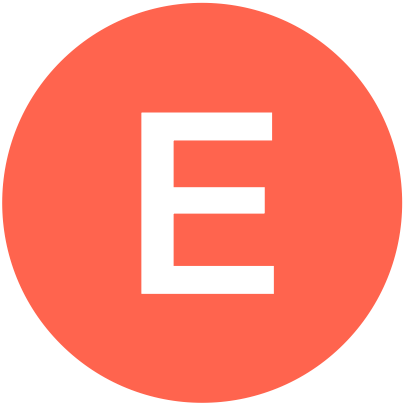}).
    %
    % \emph{SMTLIB Tutorial} (\circledletter{A}) includes tutorial content and editable code examples in SMTLIB format such as \circledletter{B}.
    % %
    % \emph{Programming Z3} (\circledletter{C}) presents Z3 materials in JavaScript and Python.
    % %
    % \emph{Playground} has a freeform editor (\circledletter{D}) and logic formula guessing games (\circledletter{E}). 
    % %
    % It also comes with links to external resources (\circledletter{F}) and a built-in search (\circledletter{G}).
    % \TOOL is a web-based environment where the user can: (A) learn about writing logical formulas using the SMTLIB language, a standard adopted by SMT solvers including Z3, and interacting with in-line examples; (B) discuss code examples, potentially with their edits, on the GitHub Discussion for Z3; (C) learn about using Z3 in languages such as JavaScript with interactive examples; (D) model symbolic logic formulas opportunistically using the freeform editor; (E) learn more about symbolic logic modeling via games; and (F) explore other resources about symbolic logic modeling and Z3.
    }
	\label{fig:overview}
\end{figure*}

\begin{figure*}[th]
	\centering
	\includegraphics[width=\textwidth]{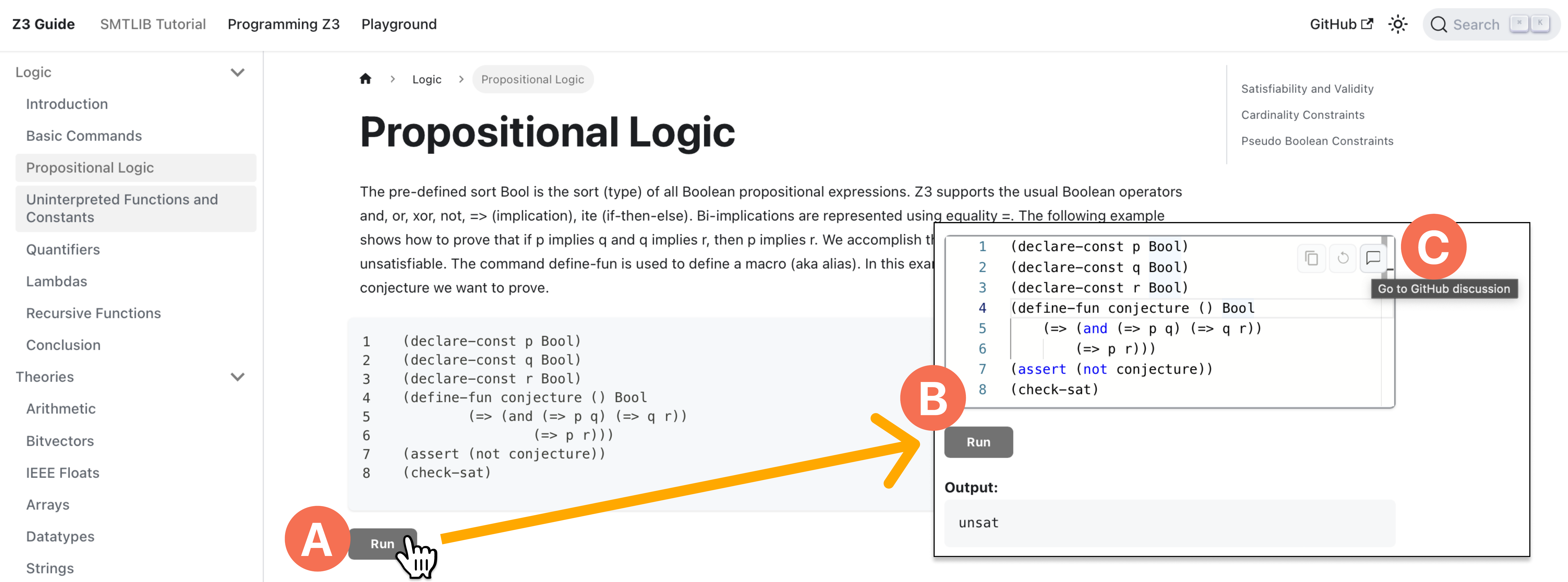}
    \Description{
    Screenshot of a section of the SMTLIB Tutorial. There are two visual columns of the content. The left column contains an outline of sections of the tutorial. The right column shows the currently selected section, Propositional Logic. The right column contains some textual description of the topic and one code example. There is an A label on the code example. The code example is written in SMTLIB format in an editor with line numbers. Below the code example, there is a Run button. Next to the code example, there is a smaller screenshot showing a different state of the code example, with a B label attached. There is an arrow that draws from the Run button below the code example in the bigger to the smaller screenshot. The smaller screenshot shows the same code example with the editor and the Run button, and additionally an Output section below the Run button showing the output of the code. At the upper right corner of the editor in the smaller screenshot, there are also three buttons with a C label attached.
    }
    \caption{
    \emph{SMTLIB Tutorial} in \tool discusses basics of logic modeling with textual explanations, practice problems, and interactive code examples in SMTLIB~\cite{smtlib} (a syntax many SMT solvers including Z3 uses). 
    \circledletter{A} Each code example is static at first.
    \circledletter{B} The user clicks ``Run'' to see code output, potentially editing and rerunning the code.
    \circledletter{C} Each editor in \tool comes with three buttons at the upper right corner, from left to right: a copy button for its content, a reset button (\reset{}) that reverts the content to original, and a discussion button that by default takes the user to the Z3  discussion on GitHub.
    }
	\label{fig:smtlib}
\end{figure*}

\begin{figure}[th]
	\centering
	\includegraphics[width=\linewidth]{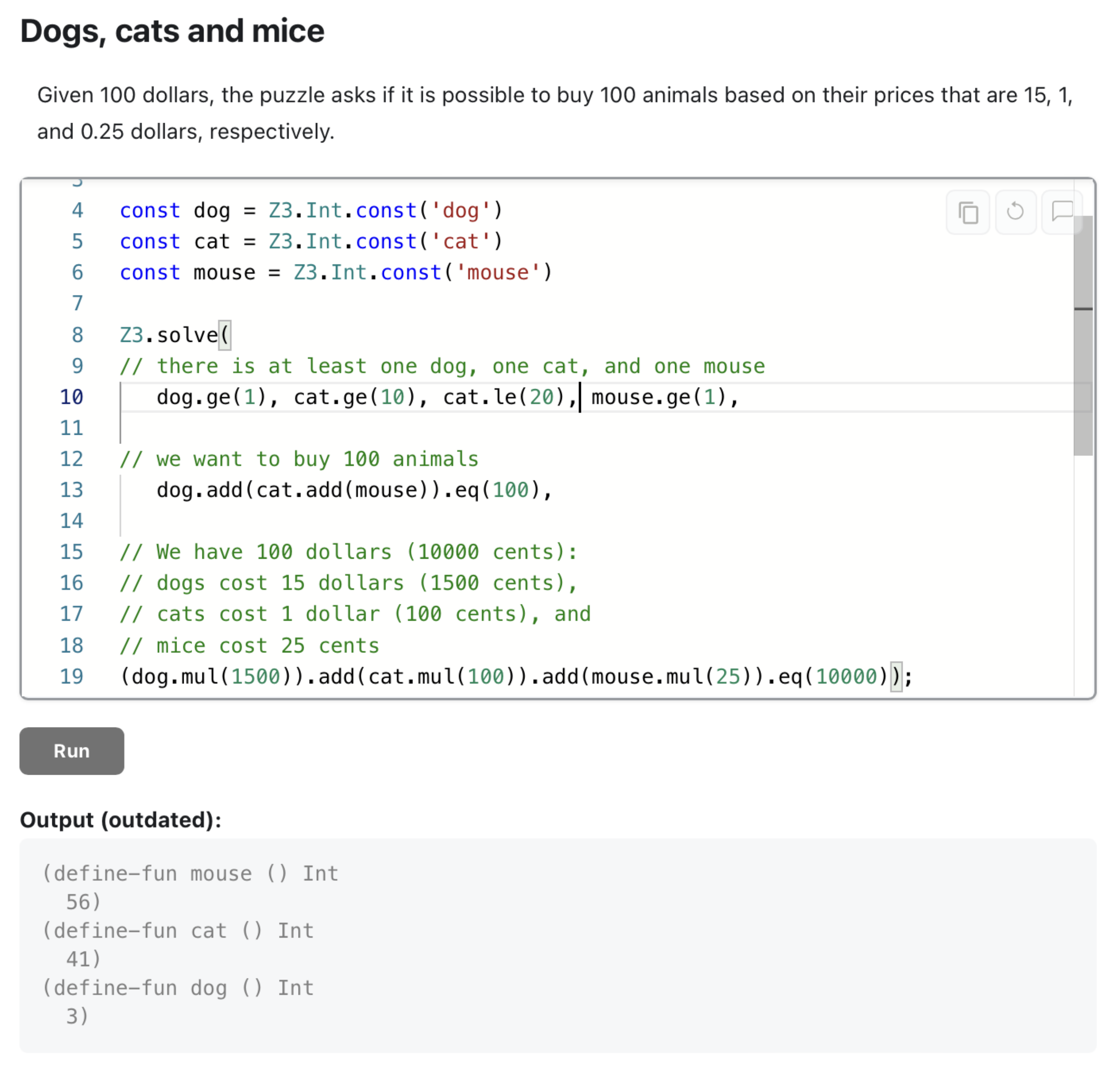}
    \Description{
    Screenshot of an example problem named Dogs, cat and mice with three visual rows. The top row shows the title and description of the problem. The middle row is an editor with JavaScript code inside that solves the problem, and a Run button below the editor. Inside the editor, the cursor is at line 10. The bottom row shows the output of the code in the editor. The output is marked as outdated.
    }
    \caption{
    An interactive logic modeling problem solved with Z3 in JavaScript. 
    The \emph{Programming Z3} section of \tool includes many of such Z3 examples as well as tutorials, introducing using Z3 in JavaScript or Python.
%    \emph{Programming Z3} has a similar format to the SMTLIB Tutorial shown in \autoref{fig:smtlib} that includes tutorials and code examples of using Z3 in Python and JavaScript.
    %
%    This  example is solved with Z3 in JavaScript.
    For every interactive example, if the user presses ``Run'' to see the output, and then edits the code, the output area fades to prompt the user to rerun the code for an up-to-date output.
%    When the user modifies the code after last running it, the output area fades to prompt the user to rerun the code.
    }
	\label{fig:js}
\end{figure}

\begin{figure}[th]
	\centering
	\includegraphics[width=\linewidth]{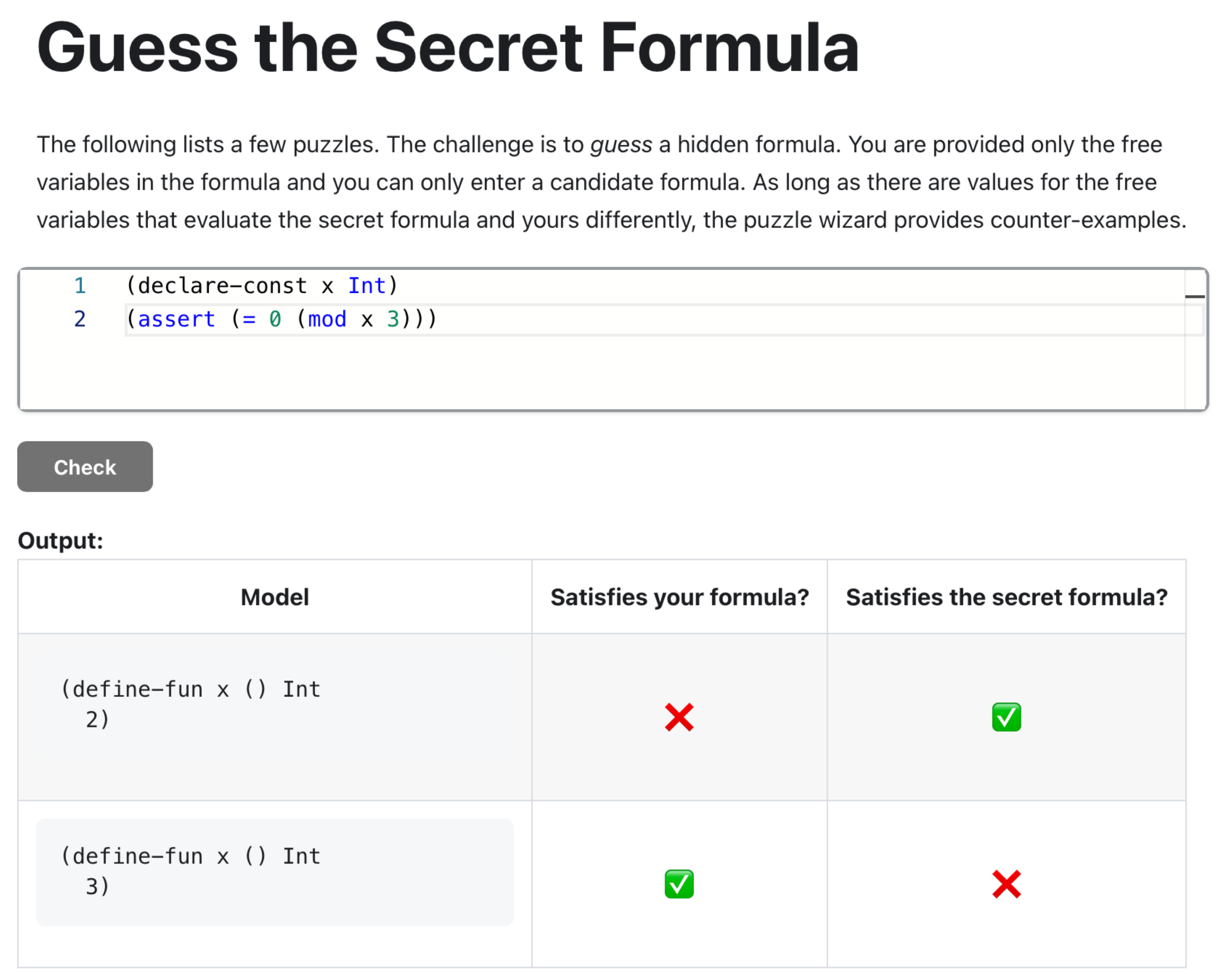}
    \Description{
    Screenshot of a section of the formula guessing games in the Playground of Z3Guide. There are three visual rows in the screenshot. The first row shows the title of the section, Guess the Secret Formula, and its description. The middle row contains an editor with some Z3 code in the SMTLIB format and a Run button below the editor. The bottom row shows the result output of the code in the editor using a table. The table has three columns, Model, Satisfies your formula?, and Satisfies the secret formula?. The table has two rows, each showing a model in SMTLIB, an emoji showing whether the model satisfies the user’s formula in the editor, and an emoji showing whether the model satisfies the secret formula.
    }
    \caption{
    \emph{Playground} in \tool contains formula guessing games.
    Each game encourages students to create a logic model that matches the secret model behind the game based on the output, which shows instances that satisfy and/or fail the user-specified model.
    }
	\label{fig:secret-formula}
\end{figure}

\begin{figure*}[th]
	\centering
	\includegraphics[width=\textwidth]{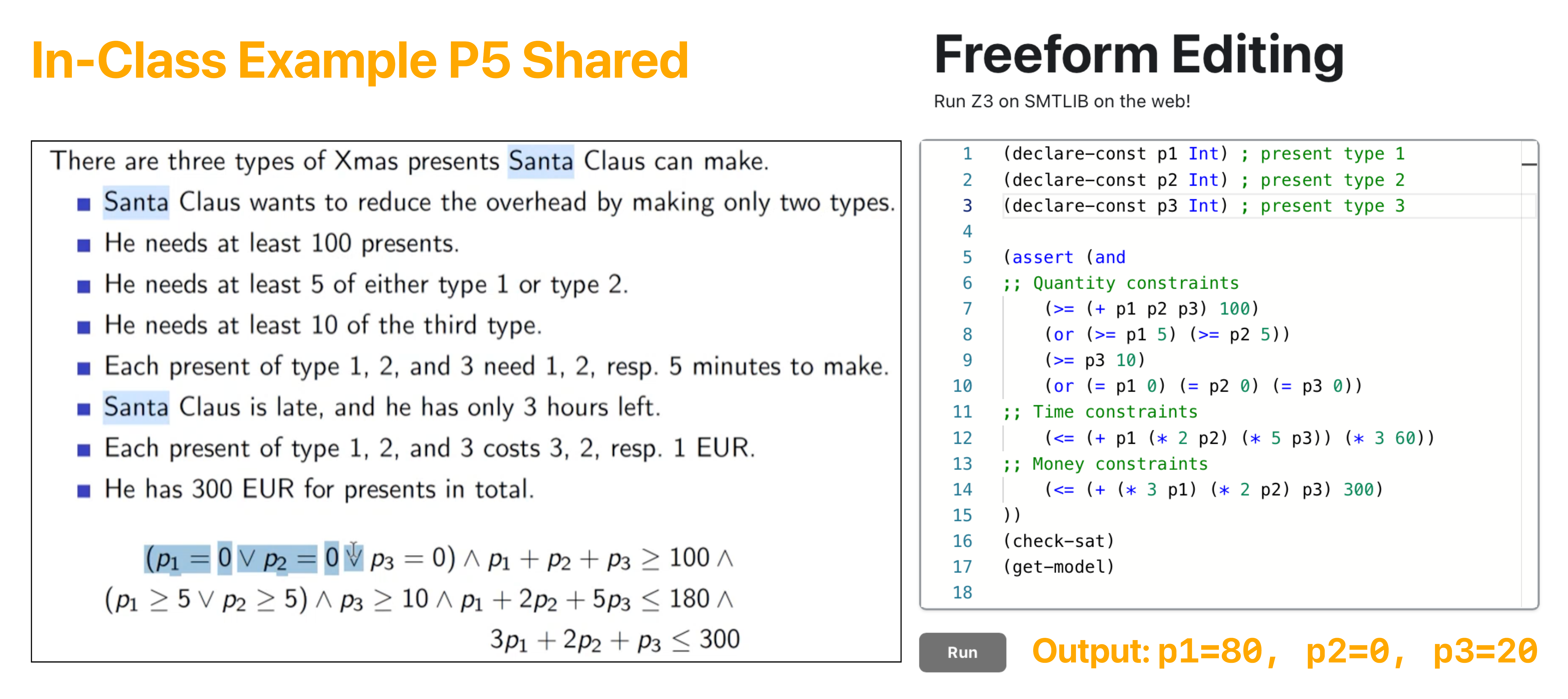}
    \Description{
    Two side-by-side screenshots. The left screenshot shows the textual description of a logic modeling problem. The right screenshot shows the freeform editor of Z3Guide with Z3 code in the SMTLIB format implementing the logic modeling problem. Above the editor is the title of the corresponding section in Z3, Freeform Editing. Below the editor is a Run button, and an annotation to the screenshot summarizing the output of the code.
    }
    \caption{
    \emph{Playground} in \tool also includes a freeform editor for Z3 in SMTLIB.
    Left: An example P5 shared in the design iteration study.
    Right: Solving the example problem in the freeform editor. The model output is summarized in orange text.
    }
	\label{fig:freeform-editor}
\end{figure*}

\subsection{Usage Scenarios}\label{subsec:usage-scenarios}
We describe two usage scenarios below to demonstrate the functionality of \tool.
% and achieves \textbf{D2: Supporting Multi-Modal Education}.
%
The first scenario is fictional based on contents in \tool, whereas the second scenario is based on how P5 used it in the design iteration (\autoref{subsec:cognitive-walkthroughs}).

\mypara{Informal Learning}
Sofia is a working professional.
Although she is not in the tech industry, she got her college degree in Applied Math and knows logic and basic programming.
As such, she is always curious about tools that help solve mathematical problems programmatically.
Recently, she heard about Z3 and logic modeling from a friend who used it to arrange seats for an event, which involves dealing with various constraints.
She decides to learn about logic modeling and Z3 in her spare time using \tool.

\begin{enumerate}[leftmargin=13pt]
    \item Sofia presses ``SMTLIB Tutorial'' (\autoref{fig:overview}-\circledletter{A}) to learn the basics of logic modeling using the Z3 SMT solver in SMTLIB format~\cite{barrett2010smt}, a syntax many SMT solvers use, following a textbook-like tutorial (\autoref{fig:smtlib}). She runs the Z3 code snippets embedded in the tutorial, sometimes with changes, to understand the concepts. When she has questions about a snippet, she presses a button inside the editor to go to the Z3 GitHub discussion page (\autoref{fig:smtlib}-\circledletter{C}) to look for related discussions on the topic or start a new discussion thread. 
    \item After Sofia acquires the basics of logic modeling and Z3, thinking about using Z3 in JavaScript, one of the languages she is most familiar with, she presses ``Programming Z3'' (\autoref{fig:overview}-\circledletter{B}) to follow tutorials on using Z3 in JavaScript. The tutorials follow the same structure as the SMTLIB tutorial with editable/executable code snippets. She finds the ``Dogs, cats and mice'' example interesting (\autoref{fig:js}), so she first runs the example to inspect its output, and then edits the code to see how the modeling results will update if she changes the quantity thresholds for cats from $cats >= 1$ to $10 <= cats <= 20$ (line 10, \autoref{fig:js}); it turns out that such constraints are unsatisfiable (not shown in figure).
    \item In the same editor, she applies the concepts behind the ``Dogs, cats and mice'' problem to budgeting for grocery shopping (not shown in figure): \emph{Given \$30 for purchasing toilet papers, fruits, and snacks, and the constraints that (1) she must buy a pack of toilet papers, which costs \$8.99 (2) she wants to buy some fruits and some snacks (3) she cannot spend more on snacks than on fruits (4) she wants to spend all the money, how should she allocate the money?} While writing Z3 JavaScript code for the problem, she explores using \scode{Real} as opposed to \scode{Int} (lines 4-6, \autoref{fig:js}) to encode constraints for budgets, since money does not need to be integers. Within seconds, the model came back to her with a solution: spend \$10.25 on snacks and \$10.75 on fruits.
    \item Sometimes, she wants more programming challenges beyond constructing logic models. She presses ``Playground'' (\autoref{fig:overview}-\circledletter{C}) to play formula guessing games (\autoref{fig:secret-formula}), which let her iterate on the logic model based on feedback from the modeling tool to guess the secret model.
    \item Sofia finds the linked materials towards the bottom of the page (\autoref{fig:overview}-\circledletter{D}) useful for her future reference, including slides for learning advanced topics of logic modeling.
    \item Throughout her usage, she finds materials for a specific concept with the built-in search feature (\autoref{fig:overview}-\circledletter{E}).
    % \item Throughout her usage, she searches among the entire tutorial site to look for materials for a specific concept.
    % \nsb{Throughout her usage, she makes use of the built-in search facility to find materials for a specific concept.}
\end{enumerate}
Using \TOOL, Sofia studies logic modeling at her own pace, with the ability to (1) read, write, edit, and run Z3 code in a variety of formats, (2) join discussions about materials in \TOOL, and (3) access other resources of more advanced concepts related to logic modeling, all without setting up a local development environment. 
Without \TOOL, she would have needed to rely on web search to look for tutorials, spend hours setting up multiple local development environments for different language bindings, and potentially sign up for workshops or courses to receive more guidance.

% \begin{enumerate}
%     \item She learns the basics of Z3 solver...
%     \item She runs code snippets... uses github...
%     \item She tries out Z3 in JavaScript and works on the dogs and cats problem
%     \item In the same editor, she explores a different problem, grocery shopping (code in comments below)
%     \item She plays secret formula guessing games to reinforce understanding of logic solving
%     \item Built-in search for finding materials
    
% \end{enumerate}

%\begin{lstlisting}[style=code, caption={Code.}, label={lst:grocery-code}]
%const fruits = Z3.Real.const('fruits')
%const sweets = Z3.Real.const('sweets')
%const toilet_paper = Z3.Real.const('toilet_paper')
%
%Z3.solve(toilet_paper.eq(4.93),
%  fruits.ge(sweets),
%  sweets.gt(1),
%  toilet_paper.add(fruits.add(sweets)).le(30))
%\end{lstlisting}

\mypara{Formal Education}
Miles is a Professor specializing in Formal Methods.
%
% Today is their first time teaching symbolic logic modeling at the university, and they are nervous.
% %
He is teaching logic modeling for 150 students in the coming school term.
This is the second time he has taught the course, and he has been thinking about possible pedagogical improvement.
Miles decides to use \tool to make the lectures more interactive and consolidate course materials.
To that end, he adds customized configuration and content to \tool.\footnote{The customization process is not shown in the figures.}

\begin{enumerate}[leftmargin=13pt]
    \item In the customized \tool, Miles changes the ``GitHub Discussion'' button inside its editor to direct students to the GitHub Discussion used by the class, as opposed to the Z3 solver discussion. From there, students could ask the class questions about a particular code snippet or concept in the tutorial. 
    \item  When preparing for teaching, he refers to slides made by other people in the community for inspiration (\autoref{fig:overview}-\circledletter{D}).
    \item Miles assigns some sections of the SMTLIB tutorial (\autoref{fig:smtlib}) as pre-class readings and exercises, including some additional examples he adds on top of the original ones. In each lecture, he discusses key concepts with the class, and, to encourage participation, changes parts of the built-in code snippets and asks students to predict the outcome.
    \item During a lecture, Miles shows on a slide a problem where Santa Claus needs to make presents given some constraints around quantity, cost, and time. Miles asks his students to pair up and model the problem in the freeform editor in \tool (\autoref{fig:freeform-editor}). All 150 students access \tool all at once. Miles later demonstrates solving the problem using the same editor himself.
    \item At the end of each lecture, he spends five minutes solving one formula guessing question (\autoref{fig:secret-formula}) with the class. His students love brainstorming solutions with one another.
\end{enumerate}
Miles bases his teaching on a customized \TOOL, aggregating all components needed for the class in one platform: readings, programming environment, and online discussions (GitHub for the class), while promoting student engagement via group work and games.
Had Miles not used \tool, he would have needed to use multiple tools for all of his pedagogical needs. 
In addition, he might have needed to troubleshoot the programming environment configuration for his students or even himself.
Most importantly, his customization needs might not have been satisfied without reimplementing some existing solutions from scratch (and knowing how to do so in the first place).

\subsection{Providing Easy Access}\label{subsec:access-and-programming}
%
% This subsection discusses how \tool addresses the first category of design guidelines \textbf{Providing Easy Access}.

\tool is web-based and completely client-side with no capacity limit, allowing hundreds of users to use it simultaneously.
This is because the Z3 formulas (in SMTLIB and JavaScript formats\footnote{While \tool includes Z3 examples in Python, they are read-only due to the lack of client-side compilation support at the time of our implementation.}) are compiled to Web Assembly (WASM)~\cite{webassembly2023} and then executed client side, without going through any external hosts.
As such, users can access \TOOL via a URL anytime, from any web browser that supports WASM, to model Z3 logical formulas and have them run directly on their machines.
The client-side implementation thus enables \textbf{D1: \emph{Fast Execution}}: when the user changes an interactive code block, the code is immediately recomputed in the browser.
%

%%
%Users can access \TOOL via a URL from any web browser that supports Web Assembly (WASM)~\cite{webassembly2023} anytime.
%% let it be a computer or a smartphone.
%%
%Without any installation or configuration, users can model Z3 logical formulas in SMTLIB or JavaScript formats because the language runtime compiles the code to WASM.
%%
%In addition, the compilation is completely client-side without going through any external hosts.
%%
%%In addition, because at build time we pre-computed the modeling results for each embedded code block, the user can see the respective output immediately if no code is edited.
%%
%In other words, if the user changes the content of a code block, the code is recompiled and the output is recomputed, delivering a fast, interactive editing experience (\textbf{D1a}).
%%
%With the client-side implementation, \TOOL has no capacity limit, allowing hundreds of users to use it simultaneously.
%%
%This setup supports \textbf{D1: Providing Easy Access}.
%%
%Note that while \tool includes Z3 examples in Python, they are read-only due to the lack of client-side compilation support at the time of our implementation.

\tool also provides \textbf{D3: \emph{Editing Support}} for programming---such as syntax highlighting and parentheses autocomplete---and additional shortcuts for copying code, undo-/redo-ing edits, and asking for help, in all editors.
When the user hovers over an editor, three buttons appear in the upper right corner (\autoref{fig:smtlib}-\circledletter{C}):
% as shortcuts for three common actions emerged in the design exploration: 
(1) a \scode{copy} button saves the current content in the editor to the clipboard for potential reuse and sharing (partially \textbf{D2} as \tool does not fully address this design goal)
(2) a \scode{reset} button resets the editor content to its original state; in case the user accidentally resets the state, the \scode{reset} button (\reset) changes to an \scode{undo} button (\undo) for three seconds to allow the user to undo the reset action and retain their work
(3) a \scode{discussion} button takes the user to some online discussion forum (the GitHub Discussion for Z3 by default) in a new tab for help with content within the guide.

\subsection{Supporting Multi-Modal Education}\label{subsec:student-centered-design}
% We achieve the second category of design guidelines \textbf{Supporting Multi-Modal Education} as follows.

\tool provides a variety of contents.
It comes with three main sections, respectively reached via pressing \autoref{fig:overview}-\circledletter{A} to \circledletter{C}: 
\emph{SMTLIB Tutorial} reviews the basics of logic and contains basic- to advanced-level topics of logic modeling with interactive examples in SMTLIB;
\emph{Programming Z3} introduces more complex Z3 program examples and API references in JavaScript (executable) and Python (read-only);
\emph{Playground} includes a freeform editor for the SMTLIB binding and a series of formula guessing games.
In particular, the formula guessing games differ from all other examples throughout \tool because they ``reverse'' the logic modeling process: while logic modeling builds a model given pre-defined constraints, the formula guessing games require changing the constraints given concrete instances of modeling successes and failures (\autoref{fig:secret-formula}) to build a satisfying model.
Such differences aim to promote a deeper understanding of logic modeling concepts.
%

%%%
With the variety of contents, \tool thus enables several use cases for the classroom.
First, via interactive \textbf{D4: \emph{Small Examples}} throughout the tutorials, \TOOL supports \emph{learning-by-doing}~\cite{anzaiTheoryLearningDoing1979}, 
% enabling educators to demonstrate concepts by running the attached examples, even with modifications, and
enabling students to internalize concepts by editing and running the examples as they go through the tutorial.
Second, with editable code blocks throughout the tool and a freeform editor in Playground, \tool not only facilitates students to perform \textbf{D5: \emph{Freeform and Exploratory Programming}} but also helps educators easily live-code~\cite{selvarajLiveCodingReview2021} in front of the class with the students following along.
This can all be done interactively, directly within the browser, because of its programming support and client-side execution (\autoref{subsec:access-and-programming}).
Third, the formula guessing games encourage learning via \textbf{D6: \emph{Gamification}} and allow educators to run engaging group activities.
Fourth, the tutorials can be supplemental materials for a course to support concept previewing and reviewing (\textbf{D7}).
Lastly, with a button that redirects to some online discussion forum (customizable) in each code example, \TOOL supports asynchronous \textbf{D8: \emph{Question Asking}} during and after class.
%effectively exposes educators and students to more available resources, online discussions, and the Z3 community.

%%%
Finally, and more importantly, all \tool contents are self-contained, allowing students to learn the topics at their own pace.

\subsection{Allowing for Extensions}\label{subsec:customization-and-extension}
%\rh{Design Goal 3}
%\begin{enumerate}
%    \item Extending/modifying existing content: create new/modified markdown documents and submit pull requests -- or host on own github pages
%    \item New language code block support:  one can configure the md code block label, line numbers, timeout for execution, language runtime npm, process for build if precomputing runtime outputs, flags for code statuses... see code example below in comments
%\end{enumerate}

% This subsection shows how \tool \textbf{allows for extensions} (third category of the design guidelines) from the user's perspective; \autoref{sec:implementation} discusses the technical implementation that enables this support.

\tool achieves \textbf{D9: \emph{Resources Sharing}} by both providing links to related Z3 resources (\autoref{fig:overview}-\circledletter{D}) and allowing contributions to its content.
\TOOL is open-source on GitHub (\textbf{D10}).%
\footnote{\toollink}
All of the tutorial contents are written in Markdown.
As such, anyone can fork the repository, edit existing Markdown files for the content or add new ones, and submit a pull request for content contribution.
%
%We configured CI/CD pipelines for automatically redeploying \tool to the web via GitHub Pages whenever a pull request is merged.
%
This allows educators to add their teaching materials to \tool for sharing with all users of the tool.

Furthermore, anyone can easily extend \tool (\textbf{D10}) as tool contributions or for their own use.
Part of its implementation enables easy extensions (which we detail in \autoref{sec:implementation}): 
(1) All contents of \tool are written in Markdown files because \tool uses docusaurus~\cite{docusaurus2023}, a static site generator, to compile the Markdown files to HTML and create its webpage 
(2) A single JSON file for \emph{configuration of languages} specifies editing and execution support for all interactive code examples.
As such, users can add interactive code examples for other Z3 bindings or other logic modeling tools by extending both the content in Markdown and the JSON configuration of languages.
%which is a JSON file that specifies additional language support for the interactive code examples.
%
For example, if one wants to add interactive examples for Dafny~\cite{leino2010dafny} (another logic modeling tool), they can (1) create Dafny code blocks in Markdown starting with \scode{\textasciigrave\textasciigrave\textasciigrave dafny} (2) in the language configuration JSON file, declare a new language named \scode{dafny} and configure support for it, such as syntax highlighting, path to the runtime for computing outputs, and destination for the Discussion button in the editor.
%
% Nikolaj: Dafny isn't an optimal example because it doesn't compile to wasm.
% There is no browser only tutorial for Dafny.
%
To enable these extensions/customizations for \tool, users can either contribute directly to the \tool repository via pull requests or deploy their fork to their own domains via mechanisms such as GitHub Pages.

\section{Implementation Challenges}\label{sec:implementation}

When implementing \tool, we realized that existing technologies could not fully address two categories of our design guidelines: \textbf{Providing Easy Access} and \textbf{Allowing for Extensions}.
This section describes the technical challenges we resolved to meet the design guidelines.
Solutions to these challenges are not specific to \tool and can apply to other interactive programming environments written entirely in Markdown.
% With solutions to these challenges, the insights from our implementation can apply to contexts beyond \tool or logic modeling, such as an interactive textbook for Python written entirely in Markdown.

%All contents of \tool are written in Markdown.
%%
%\tool uses docusaurus~\cite{docusaurus2023}, a static site generator, for compiling the Markdown files to HTML and packaging the website.
%%
%In addition, it uses a Z3-to-Web Assembly compiler for computing the Z3 programs client side.
%%
%However, there are two challenges, which using existing tools cannot resolve, per our design goals \textbf{D1} and \textbf{D3}.
%%
%First, Markdown is used for generating read-only content, and the syntax highlighting supports a limited set of languages; our tool should allow contributors to create interactive code examples with proper highlighting for various Z3 formats (e.g., SMTLIB and JavaScript).
%%
%Second, some larger Z3 examples in the tutorial might take a while to be computed at runtime; these examples should not be removed as they could be important for learning, while our tool should still deliver an interactive experience that responds to the user in seconds.

% https://docs.github.com/en/get-started/writing-on-github/working-with-advanced-formatting/creating-and-highlighting-code-blocks

%%%
\subsection{Interactive Code Examples in Markdown}\label{subsec:interactive-code-examples}
%
% \mypara{Intended Use Case}
Our design guidelines state that anyone, particularly those who teach logic modeling, should be able to easily extend \tool (\textbf{D9}-\textbf{D10}), including (1) writing new Z3 examples that will be rendered as interactive code snippets and (2) adding examples in other language bindings for Z3 or other logic modeling tools, with proper syntax highlighting and runtime for execution.
However, all \tool contents are written in Markdown, which 
% \mypara{Limitations with Existing Techniques}
% %
% All contents of \tool are written in Markdown.
% %
% However, Markdown 
by default renders static content only, not editable content, and the syntax highlighting for code blocks in Markdown only supports a limited set of languages.
%
%our tool should allow contributors to create interactive code examples with proper highlighting for various Z3 formats (e.g., SMTLIB and JavaScript).

% \mypara{Our Solution}
%
We contribute three techniques to support the intended use case: 
(1) a customized editor (as shown in \autoref{fig:js}) that provides proper syntax highlighting and autocomplete for a Z3 code example, the buttons for the copy, reset, and discussion actions, an output area that shows the result of execution, and a ``Run'' button that triggers computing the output;
(2) a Markdown rendering plugin\footnote{https://github.com/remarkjs/remark} that replaces the default read-only code blocks with interactive code blocks with our customized editor and computes the outputs of the code blocks when applicable;
(3) a JSON file---configuration of the languages (denoted as \scode{lang-config} below)---that specifies code blocks of which languages should be replaced with the customized editor, and additional language-specific information for the rendering and output computation.

We implemented these techniques for the Z3-SMTLIB and Z3-JavaScript examples in \tool.
\autoref{lst:lang-config-code} shows part of the configuration for the Z3-SMTLIB code blocks in \scode{lang-config}.
% including the label for the Markdown code blocks (line 3), syntax highlighting to be used (line 4), destination of the discussion button (line 14)
%
Using \scode{lang-config}, for each code block in Markdown, the plugin properly specifies the syntax highlighting (line 4), the process for computing the code (lines 10-11) during build (\autoref{subsec:computing-tutorial-examples}) or at runtime (associated with the ``Run'' button), the destination for the discussion button (line 14), etc. for each instance of the customized editor.
The plugin further computes the output of the code (more in \autoref{subsec:computing-tutorial-examples}) if the \scode{buildConfig} property is specified (lines 6-13, \autoref{lst:lang-config-code}).

Note that while \tool is 100\% client-side for Z3, users could still implement output computation that relies on external servers when extending \tool with examples in other languages.
They could write a script that handles the client-server communication and set \scode{processToExecute} (line 10, \autoref{lst:lang-config-code}) to the script.

Because our approach combines Markdown rendering and configuration for specific language support, it is intended to generalize to other Markdown-based interactive environments. 
For example, tutorials for MSAGL\footnote{\url{https://microsoft.github.io/msagljs/}} are built with the framework behind \tool.

\begin{lstlisting}[
	style=code, 
	caption={Configuration of Z3-SMTLIB blocks}, 
	label={lst:lang-config-code}
]
{
	name: 'Z3', // your language name
	label: 'z3', // label for markdown code blocks
	highlight: 'clojure', // prism-supported syntax highlighting
	showLineNumbers: true, // whether to show line numbers
	buildConfig: {
		timeout: 30000, // execution timeout of each snippet in ms
		npmPackage: 'z3-solver', 
			// npm package name for the language runtime, if any
		processToExecute: './src/remark/run-z3-smtlib.js', 
			// process for computing code outputs
		...
	},
	githubRepo: 'Z3Prover/z3', // discussion button destination
	githubDiscussion: true // whether to show the discussion btn
}
\end{lstlisting}

\subsection{Computing Tutorial Examples}\label{subsec:computing-tutorial-examples}

% \mypara{Intended Use Case}
Our design guidelines also state that anyone should be able to easily access the tool and its interactive code examples (\textbf{D1}-\textbf{D3}).
% : even large examples in the tutorials might take long to compute, users should not have to wait too long.
% \mypara{Limitations with Existing Techniques}
%
Since the static site generator used in \tool, docusaurus~\cite{docusaurus2023}, does not handle Z3 execution directly, we initially attached a Z3-to-WASM compiler to \tool to compute Z3 examples client-side when the user runs the examples.
% our initial approach to turn static Z3 examples interactive was to simply attach a Z3-to-WASM compiler to \tool that computes Z3 examples client-side when the user runs the examples.
%
However, larger Z3 examples might still take long to be computed at runtime, and the user's computing environment can affect the client-side computation efficiency.
We deem these large examples essential as they could be important for learning, but \tool should still deliver an interactive experience that responds to the user in seconds.

% \mypara{Our Solution}
%
We implemented three approaches to ensure an interactive and responsive experience with \tool in as many cases as possible.
% (1) precomputing code outputs when building \tool (2) caching the precomputed outputs to avoid unnecessary re-computations during build (3) configuring a timeout or user-initiated executions at runtime.
First, we precompute all code examples in \tool when building from the source (prior to deployment), including the large examples that might be slow to compute.
Second, we cache the precomputed outputs by writing each code output to a file on the disk and naming it with a hash of the code, the language runtime name, and the runtime version.
This way, we avoid unnecessary re-computations during the build, only recomputing the output of an example if it is new or the language runtime is updated.
Third, we configure the timeout of the runtime execution (\autoref{lst:lang-config-code}) to be 30000 milliseconds (30 seconds): if the user edits a large example and recomputes it at runtime, the execution will timeout after 30 seconds and prevent them from waiting too long.
Most examples in the tutorial will not reach this timeout threshold, and if the user encounters an unexpected execution timeout or wants to get some long example to be computed, they can ask for help in the discussion.

%
%%
%However, while this decision saves users some time when running large tutorial examples for the first time, it does not help when they edit a large example and recompute it at runtime.
%%
%We configured the timeout of the runtime execution (\autoref{lst:lang-config-code}) to be 30 seconds to prevent users from waiting too long.
%%
%This threshold should apply to most examples in the tutorial, and if a user encounters an unexpected execution timeout or wants to get some long example to be computed, they can ask for help in the discussion.

With this approach, failures in precomputing code examples at build time will cause the build to fail and thus help contributors catch examples with unintentional errors.
However, contributors might want to create examples with \emph{intentional} errors for educational purposes.
They could skip computing these examples at the build by adding a \scode{no-build} flag when creating the Markdown code blocks (e.g., \scode{\backtick\backtick\backtick z3 no-build}).
Our Markdown rendering plugin (\autoref{subsec:interactive-code-examples}) will thus skip the computation of examples with this flag.

\section{Students Using \tool}\label{sec:deployment}
%\red{reconsider the justification for evaluating on students only not educators}
%
%\red{should also emphasize that all participants are \emph{students}}
%
% Despite a design iterated with \emph{educators}, 
The design of \TOOL was driven by needs and feedback from the \emph{teachers}.
To understand how \emph{students} perceive \TOOL and gauge possible improvements for the tool, we conducted a free online workshop where participants used \TOOL to learn logic modeling and Z3.
% and provided their feedback through a survey. 
%
We advertised the workshop through mailing lists of our institutions and on social media, eventually getting 112 attendees.
We informed them of the workshop agenda and its research purpose and asked for consent to data collection during registration.
%
% Participants were not compensated, but the participation in the workshop was free.
% \nsb{This doesn't sound right, the "but" does not contrast the previous part of the sentence. Maybe say:  Participants were not compensated. We offered a free workshop in lieu of compensation.}
Only 21 attendees gave consent, and per our IRB protocol, we only used their survey responses to report findings.

\mypara{Participants}
We recruited 21 adult participants (4 women, 16 men, and 1 unknown) from 11 countries and regions.
% All except three were full-time working professionals in computing-related fields.
%
All participants are \emph{novices} to logic modeling and Z3: 19 participants had no experience in logic modeling or Z3, and two participants self-reported minimal prior experience.
% Most participants had no experience in logic modeling or Z3, except for two with limited prior experience.
Four participants were pursuing a Bachelor's degree in Computer Science, while the rest were full-time working professionals in computing-related fields.
All participants joined the workshop because of interest in the topic and/or using the technique in their work.
% who had or were pursuing a Bachelor's degree.
%
Six belonged to the 18-24 age group, and two were above 60.
%
%
% Our research protocol was approved by the institution's Review Board.

\mypara{Procedure}
%
%\red{What did the participants do? How did they use the system? What topics were covered?}
%
%\red{Emphasize the interactive aspect of the workshop -- teacher shows how to use the tool, other authors helping out in the chat for troubleshooting, students interacting with the teacher and fellow students in real time in the chat / by unmuting themselves}
%
We conducted the workshop via web conferencing.
The workshop lasted three hours:
% covering a variety of topics in logic modeling with Z3: 
two 75-minute tutorials with a 15-minute break in between, and 15 minutes at the end for a post-workshop survey.
The first tutorial covered the basics of logic modeling with Z3 in the SMTLIB format following parts of the SMTLIB Tutorial in \tool (\autoref{fig:smtlib}) and the formula guessing games (\autoref{fig:secret-formula}), while the second introduced using the Z3 solver API in JavaScript and Python for solving realistic constraint-satisfaction problems (e.g., \autoref{fig:js}).
Only \tool
% , along with its conceptual explanations and interactive examples for logic modeling and Z3, 
was used in the workshop, with no other supplementary materials, although participants were free to search for related content on the internet.
All workshop attendees, including those excluded from the analysis, used \tool simultaneously while following the tutorials and programming within the environment.
The last author led the workshop, lecturing and facilitating discussions and exercises.
Workshop participants interacted with one another by unmuting themselves in the web conference or messaging within the chat.
% providing feedback on \TOOL in a survey.
%
Three authors helped moderate the workshop, including admitting attendees, monitoring the chat, and troubleshooting \tool when necessary.

\mypara{Post-Workshop Survey}
At the end of the workshop, participants filled out a survey reflecting on their experience using \tool.
The survey consisted of five Likert-scale rating questions about the programming experience within \tool, its overall user experience, and whether participants would recommend the tool to other similar users; \autoref{tab:rating} summarizes the statements used in the questions.
The survey also included three short response questions asking the participants about features they liked, disliked, and wished to be available in \tool.
In addition, participants provided demographic information in the survey.

\mypara{Data Collection and Analysis}
We monitored comments in the chat related to issues of \tool, though we did not record any.
In addition, we collected post-workshop survey responses, which included demographics,~Likert-scale~ratings,~and open-ended comments on \TOOL.
% All data collected from the need-finding interviews, the prototype evaluation sessions, and the post-workshop survey are qualitative.
%
The first author coded the comments using open coding
% following the thematic analysis guidelines
~\cite{braunUsingThematicAnalysis2006}.
Using the same approach in~\autoref{subsec:need-finding-interviews}, the first two authors computed the inter-coder agreement using 25\% of the data and achieved a 92\% agreement.

\mypara{Limitations}
%%% WORKSHOP LIMITATIONS
%
The deployment only evaluated \tool in one aspect---how it could affect students' learning experience---while there are many other aspects of \tool and the design guidelines behind to be more thoroughly evaluated, such as its impact on learning outcomes, the teachers' experience, and the generalizability of its implementation to other educational tools for logic modeling.
Our evaluation is also limited to its sample size, with only a fifth of the 112 attendees giving consent to using their post-workshop survey responses.
% %
% We also only collected post-workshop survey responses and did not probe how students used \tool for learning during the workshop.
%beyond recording their perceptions and feedback.
%
%As such, our deployment was exploratory in nature and a formative evaluation of \tool.
% the design of \TOOL.
% As such, our deployment was exploratory in nature and aimed at analyzing qualitative feedback on \TOOL.
%
As such, the deployment should be viewed as an early step towards understanding the use of \TOOL and its design guidelines, future summative assessments, and future tool designs in related domains.
%As such, Our study should be viewed as an early step towards understanding the use of \TOOL, future evaluations in a classroom setting, and future tool improvements and extensions.

%%
%\todo{cohen's kappa}
%%
%Finally, the coder derived themes from the codes for each dataset.

%%%%%%%%%%%%%%%%%%%%%%%%%%%%%%%%%%%%%%%%%%%%%%%%%%%%%%%%%%%%%%%%%%%%%%
\subsection{Post-Workshop Survey Results}\label{sec:deployment-takeaways}

The workshop, with more than 100 people using \tool simultaneously, went smoothly with no reported technical issues.
Below we report findings from 21 post-workshop survey responses.
% around two themes emerged from open-ended survey comments.
% : \tool's usability and helpfulness for learning.
%
% Overall, while participants identified issues to be further improved in \tool, they found \tool and its programming support to be easy to use and helpful for learning logic modeling.

% We report findings from 21 post-workshop surveys regarding \TOOL's usability and helpfulness for learning.
% \autoref{tab:rating} summarizes the Likert-scale ratings.
% % regarding usability and helpfulness of \TOOL for learning.
% %
% %
% %Overall, participants would strongly recommend \TOOL to other users alike ($avg.=4.71, sd=.46$).
% %
% %This section reports the results and implications.

\mypara{Participants Deemed \tool Easy to Use}
On average, participants rated 4.43 ($sd=.60$) \footnote{$sd$ - standard deviation.} out of 5 (Strongly Agree) to the statement ``\TOOL was easy to use''.
Participants appreciated the ease of accessing everything used in the workshop within one tool, from tutorial materials to the programming environment, and that everything 
\emph{``loaded up instantaneously.''}
They considered the content to be \emph{``very structured,''} the examples to be \emph{``clear,''} and \emph{``there is a connection between most topics in the tutorial.''}
They found the tool to have promising usability \emph{``considering students in different levels.''}
%
% One participant worte, ``\TOOL is awesome! Tons of material and is interactive.''
%
They also strongly agreed that ``programming in \TOOL was easy'' (4.76 out of 5, $sd=.54$).
% As for \TOOL's programming experience, participants gave an average rating of 4.76 out of 5 ($sd=.54$).
%
They believed that it was easy to create and modify code examples to see \emph{``different [Z3] features in context.''}
In addition, it is \emph{``a great [tool] for playing with Z3,''} giving users \emph{``the ability to try out [Z3] without installing [it] locally.''} 

\mypara{Participants Found \tool Helpful for Learning}
Participants gave an average rating of 4.57 out of 5 ($sd=.60$) that ``\tool was helpful for learning,'' and an average of 4.71 out of 5 ($sd=.46$) that they ``would recommend \tool to other students.''
Participants commented that the inclusion and depth of \emph{``a lot of important concepts [about logic]''} could \emph{``apply to everyone,''} whether or not they intend to use Z3 for logic modeling, which provide \emph{``theoretical reminders all along the way''} while learning new concepts.
One participant noted that the tutorials went a bit too fast for them, but: \emph{``now I can read the pages in my own pace, which is really great.''}
Programming in \TOOL was considered as particularly helpful for learning ($avg.=4.76, sd=.54$): seeing the output of each snippet helped the participants refine \emph{``[their] understanding of the material and the examples.''}
Furthermore, participants liked the editable examples throughout the guide as they were \emph{``easy to run''}, with fast and stable execution, and encouraged learning-by-doing.
%
% Apart from its interactivity, some participants attributed the guide's help for learning to its accessibility: there were editable examples throughout the guide that were ``easy to run'' and encouraged example-based learning, and the execution was fast and stable.

\mypara{Participants Wanted More Programming Support and Engagement}
Several participants pointed out issues with the programming experience in \tool.
One participant, new to Z3, was frustrated about not being able to \emph{``get the syntax right''} as the editor had limited support in autocompleting keywords.
Some participants hoped for more readable, \emph{``pretty-printed''} modeling output messages that are usually long in logic modeling.
Additionally, some participants hoped for more learning engagement.
One of the highlights of \tool is logic puzzles, with one dedicated section of formula guessing games in its Playground and realistic puzzles like Sudoku throughout the tutorial.
Participants deemed the puzzles important for an engaging self-paced learning experience of logic modeling but expected more such \emph{``guided modeling problem[s].''}
They further yearned for the ability to share their work-in-progress logic modeling code with others for help, such as \emph{``via permalinks,''} rather than having to manually copy and paste the examples.

% It is not uncommon to face such tradeoffs in human-centered design as we attempt to balance user needs and system limitations.
%
% NSB: this sentence seems to
% state something obvious, so I comment it out:
%We leave it to future work for a more secure way of sharing learning progress for a better experience.

% Table generated by Excel2LaTeX from sheet 'Likert Scale'
\begin{table}[t]
  \centering
  \small
  \Description{
  Table with six rows, the first row as the header row, and four columns, the first column as the header column. The last column contains five bar charts for each of the five rows. Each bar chart shows the distribution of Likert-scale ratings corresponding to each statement presented in the header column of the respective row, with the Y axis showing the counts for each rating from 1 to 5 on the X axis. All bar charts have the bar on the rating of 5 (Strongly Agree) being the highest bar, followed by the bar on the rating of 4 (Agree) being the second highest bar. We first show the table data, then the data for each bar chart.
    The table is as follows from left to right:
    (No Text) Avg. Mdn. Dist.
    Programming in Z3Guide was easy. 4.76 5.0 (Bar chart 1)
    Programming in Z3Guide was helpful for learning. 4.76 5.0 (Bar chart 2)
    Z3Guide was easy to use. 4.43 4.0 (Bar chart 3)
    Z3Guide was helpful for learning. 4.57 5.0 (Bar chart 4)
    Would recommend Z3Guide to other students. 4.71 5.0 (Bar chart 5)
    Bar chart 1 shows the distribution of Likert-scale ratings corresponding to the statement Programming in Z3Guide was easy. There are three bars. The data is as follows:
    Rating Count
    1 0
    2 0
    3 1
    4 3
    5 17
    Bar chart 2 shows the distribution of Likert-scale ratings corresponding to the statement Programming in Z3Guide was helpful for learning. There are three bars. The data is as follows:
    Rating Count
    1 0
    2 0
    3 1
    4 3
    5 17
    Bar chart 3 shows the distribution of Likert-scale ratings corresponding to the statement Z3Guide was easy to use. There are three bars. The data is as follows:
    Rating Count
    1 0
    2 0
    3 1
    4 10
    5 10
    Bar chart 4 shows the distribution of Likert-scale ratings corresponding to the statement Z3Guide was helpful for learning. There are three bars. The data is as follows:
    Rating Count
    1 0
    2 0
    3 1
    4 7
    5 13
    Bar chart 5 shows the distribution of Likert-scale ratings corresponding to the statement Would recommend Z3Guide to other students. There are three bars. The data is as follows:
    Rating Count
    1 0
    2 0
    3 0
    4 6
    5 15
  }
  % \vspace{1em}
  \caption{\textbf{Likert-scale ratings on \TOOL in the post-workshop survey.} 1 - ``Strongly Disagree'', 5 - ``Strongly Agree''. 
  Avg. - Average, Mdn. - Median, Dist. - Distribution.
  %
  % ``Interactivity'' refers to the ability to continuously edit and run code snippets and see their results.
  }
    \begin{tabular}{lccc}
    \toprule
          & \multicolumn{1}{c}{Avg.} & \multicolumn{1}{c}{Mdn.} & \multicolumn{1}{c}{Dist.} \\
    \midrule
    \TOOL was easy to use. & 4.43  & 4.0   & \raisebox{-.4em}{\includegraphics[height=1.6em]{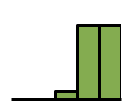}} \\
    Programming in \TOOL was easy. & 4.76  & 5.0   & \raisebox{-.4em}{\includegraphics[height=1.6em]{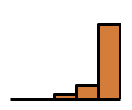}} \\
    \TOOL was helpful for learning. & 4.57  & 5.0   & \raisebox{-.4em}{\includegraphics[height=1.6em]{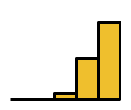}} \\
    Programming in \TOOL was helpful for learning. & 4.76  & 5.0   & \raisebox{-.4em}{\includegraphics[height=1.6em]{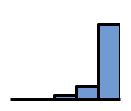}} \\
    Would recommend \TOOL to other students. & 4.71  & 5.0   & \raisebox{-.4em}{\includegraphics[height=1.6em]{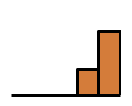}} \\
    \bottomrule
    \end{tabular}%
  \label{tab:rating}%
\end{table}%

\section{Discussion and Future Work}\label{sec:discussion}

%In this section, we revisit the questions we posed in~\autoref{sec:intro}: 
%\emph{How does one build an educational environment for logic modeling that meets the needs of the educators and the students?} (\autoref{subsec:design-reflection})
%\emph{When designing such a system, how does one balance user experience, maintenance effort, and scalability?} (\autoref{subsec:heuristic-analysis})
%We further reflect on one author's experience using \tool for a year (\autoref{subsec:author-evaluation}).
%% and the limitations of our work (\autoref{subsec:limitations}) 
%Combining these implications, we discuss future design opportunities.

We reflect on our design guidelines (\autoref{subsec:design-reflection}) and the last author's experience using \tool (\autoref{subsec:author-evaluation}) to discuss the implications of our work and future design opportunities.
%
% We further relate our design to existing design guidelines for scalable software (\autoref{subsec:heuristic-analysis}) to derive insights for designers of future educational tools.

% \red{How people perceive and use the system, implications for HCI systems research, particularly educational technologies}
% \begin{enumerate}
% 	\item user expectations for programming experience in this tool are high -- almost IDE like
% 	\item expectations for collaborative learning
% 	\item implications of human-centered design for future educational technologies
% 	\item author evaluation
% 	\item limitations
% \end{enumerate}

\subsection{Reflections on Design Guidelines}\label{subsec:design-reflection}

% Towards better educational environments for logic modeling and programming education, our design exploration with educators (\autoref{sec:design-exploration}) and evaluation workshop with students (\autoref{sec:deployment}) imply the following future work directions.

%Through a two-step design exploration with educators in logic modeling (\autoref{sec:design-exploration}), we studied their needs for an educational environment and the potential needs of the students based on their experience.
%%
%We surveyed students after they used \tool to learn logic modeling in a workshop to both evaluate their perceptions and surface unaddressed needs.
%%
%This entire design process implies the following possible improvements to \tool.
%%

Student feedback from the formative workshop suggested three design requirements missing in \tool: one was \textbf{D2: Code Sharing} that we did not implement, and the other two were possible new guidelines.
Further assessments of \textbf{D6: Gamification} also remain: while students and teachers showed a positive attitude, prior work reveals unresolved debates around its impact on student learning.

\mypara{Sharing of Learning Progress}
The post-workshop survey shows that students hoped to share learning progress, particularly via permalinks.
Indeed, code sharing (\textbf{D2}) is the only design guideline we did not fully address despite the request of teachers.
We did not implement code sharing in \tool due to security concerns: 
a permalink includes complex parameters, such as code to be parsed and evaluated, which must be done with care to prevent the possibility of malicious attacks especially when there is JavaScript execution of user-specified code using \scode{eval()}, in our case the Z3 code in JavaScript format.
%
% code would be decoded from a given permalink for evaluation, but \TOOL includes JavaScript code evaluated using \texttt{eval()}.
% , which would expose the user to code exploitation if permalinks were available. 
%
In future work, we aim to improve how JavaScript code is parsed and executed to better support user needs.

\mypara{Full Programming Support Even in a Lightweight Environment}
In the post-workshop survey, students expected \TOOL
% , despite a lightweight educational tool, 
to incorporate similar programming support to existing IDEs, such as keyword autocomplete and API discovery.
Such expectations might relate to the fact that most participants were casual learners experienced in computing, possibly with prior exposure to professional programming tools.
Design tradeoffs remain between enabling a programming experience with full IDE support and delivering a lightweight educational environment that covers other experiences than programming.
Our work prioritizes the latter but continues to seek better editing support in future iterations.
%
%In the survey, students expressed their preference for interactive code examples in the tutorial materials.
%%
%However, students also expected \TOOL
%% , despite a lightweight educational tool, 
%to incorporate similar programming support to existing IDEs, such as keyword autocomplete and API discovery.
%%
%Such expectations might relate to the fact that most participants were casual learners with prior exposure to modern programming tools given their expertise in computing (\autoref{sec:deployment}). 
%%
%There remain design tradeoffs between enabling a programming experience with full-fledged IDE support and delivering a lightweight educational environment that covers other experiences than programming.
%%
%Our work prioritizes the latter, which we discuss more in~\autoref{subsec:heuristic-analysis}.

\mypara{Comprehension Aids for Logic Modeling Outputs}
Both P4 and P6 in the design exploration and students in the workshop wanted better aids for understanding the output of logic modeling.
There is a long line of research in the output representations of algorithmic programs that facilitate code comprehension.
Even if logic modeling differs from algorithmic programming, users share a similar desire to understand code behavior via some aids.
Designing better representations for logic modeling outputs, however, has been an open problem~\cite{goualard1999visualization} and warrants a separate investigation.
Future work could leverage program visualization techniques such as in-situ visualizations~\cite{hoffswellAugmentingCodeSitu2018, huang2024unfold} to better connect the non-algorithmic logic-modeling programs with their outputs.
%

%Designers of future educational tools should take into account the target educational settings and user profiles.

\mypara{Understanding of the Impact of Gamification on Student Learning}
Both our design exploration (\autoref{sec:design-goals-of-tool}) and post-workshop survey reveal a positive attitude towards the educational games in \TOOL, while there have been debates on the use of gamification in education~\cite{deci1999meta, caponetto2014gamification}.
Future work should evaluate \TOOL in formal educational settings, particularly its gamification, to understand its impact on student learning.

\subsection{Additional Roles for Interactive Textbooks}\label{subsec:author-evaluation}

From the last author's year-long experience in using \tool to answer questions about SMT solvers in an online forum, we identify two previously under-explored roles beyond an educational environment for \tool and similar tools with interactive textbooks.
%
%
%One author has been using \TOOL to answer questions about SMT in an online forum for a year since its deployment.
%%
%Anecdotally, we report their experience in using \TOOL,
%% for such informal learning contexts, 
%from which they identify opportunities for \TOOL to be more than an interactive textbook.

% In addition to introducing users to logic modeling in classroom settings, \TOOL is integral to several aspects of documenting the Z3 solver. 
\mypara{\TOOL As A Reference Guide}
\tool is very suitable for a quick reference for logic modeling and Z3 because of its interactive examples.
Indeed, the author has been referring to \tool when answering user questions and demonstrating Z3 features.
To this end, they added new examples to \tool when there were no suitable ones or when new Z3 features came out.
Such benefits are not special to \tool or logic modeling; in fact, any extensible interactive textbooks for programming tools would be suitable for demonstrating the use and features of the tools.

%% The author considers \TOOL to be not only a learning guide but a reference to logic modeling and Z3.
%%
%% They have 
%The author has referred to \TOOL when answering user questions and added new examples to it when no existing examples covered such questions.
%% It is used as a reference when answering user questions, and new examples are added when there are no existing examples that cover such questions. 
%They have also been updating \TOOL whenever new Z3 features come out.
%% It is used when documenting new functionality and it connects to summaries of configuration parameters and exposed functionality. 
%% Thus, \TOOL is not only a learning guide but a reference to logic modeling and Z3.
%% Thus, it meshes the purpose of a learning guide with the purpose of a reference. 
%Central to enabling the reference role of \TOOL is the ability to run examples within the browser.
%% Central to making the reference useful is the integration with examples that can be run from the browser. 
%For instance, Z3 exposes a set of \emph{tactics}, which represent specialized solving steps (e.g., a formula simplification) that a user can compose to create a custom solver. 
%% A tactic represents a specialized solving step, such a formula simplification, a case split, or a transformation of a formula into a format that is more suitable for a specialized back-end. 
%For each tactic, \TOOL provides not only static documentation but also an executable and modifiable example to illustrate its use.
%%

\mypara{\tool As A Personalized Tutor}
%\tool cannot answer questions directly yet, but it is possible to enable this ability.
%%
While one can obtain answers to many questions of logic modeling or Z3 by having a human expert direct them to respective sections in \tool or searching within \tool themselves, \tool itself cannot answer ``how-to'' questions sufficiently.
%% The author also sees opportunities towards a more active role in question-answering for \TOOL and interactive textbooks in general.
%%
%% A lot of questions posted in the online forum look for answers with an example, and the tutorial format in \TOOL is already suitable for such questions.
%% One can already answer many questions posted in the online forum by reusing examples from \TOOL.
%Specifically, while one can already obtain answers for many questions by having a human expert in logic modeling and Z3 directing them to respective sections in \TOOL or searching within \TOOL themselves.
%%
%% answers to ``how-to'' questions are more complex and often involve a sequence of steps, each of which leads to specific sections in \TOOL, and a human expert is necessary to help break the solution down to steps in the first place.
%However, there are ``how-to'' questions with which \TOOL alone cannot provide sufficient help.
%
For example, to answer
\emph{``Is it possible to dump an SMT2 file after  simplifications? If so, how?''},
one needs to (1) break the solution down to steps and (2) connect each step to sections in \TOOL.
%
%
% For a more complex ``how-to'' question, the answer would involve a sequence of steps, each of which leads to specific sections in \TOOL.
%
% For example, to answer
% % A sample question that meshes well with the tutorial format is 
% \emph{``Is it possible to dump an SMT2 file after  simplifications? If so, how?''}
% one needs to be an expert who knows (1) how to break the solution down to steps and (2) where in \TOOL each step can be referenced to.
% Still, an expert is needed to break the workflow into smaller steps to guide users to respective sections in the tutorial.
%
Currently, \TOOL directly assists with step (2) via built-in search, but a human expert is necessary for step (1).
% Currently, \TOOL assists users with step (2) with built-in content search, but a human expert in logic modeling and Z3 is necessary for step (1), with which is where students need the most help.
%
Large language models (LLMs) are, on the other hand, very suitable for automating constructing the how-to solution with decomposed steps~\cite{yao2023tree} and hence reducing the barriers to obtaining answers.
We imagine \TOOL and similar interactive textbooks, which already engage students more than static e-textbooks~\cite{pollari2017value}, to become even more active in providing quick answers and feedback to students with the help of LLMs.
% We imagine question answering within \TOOL and interactive textbooks alike to be more accessible with LLMs: students can receive an answer at no delay, and educators/experts can improve the quality of LLM-generated answers with their expertise. 
%
Future interactive textbooks could support learning new concepts and receiving timely feedback on learning via integration with LLMs.

\section{Conclusion}

To understand the pedagogical needs for educational tools for logic modeling, we conducted a needfinding interview and a design iteration with teachers of the subject.
% We conducted a needfinding interview and a design iteration with teachers of logic modeling to understand their pedagogical needs for educational tools for this subject.
%
Driven by 10 design guidelines from the design exploration, we developed \tool, a scalable, student-centered, and extensible educational environment for logic modeling with the Z3 SMT solver.
We ran a workshop with more than 100 students using \tool to learn about logic modeling with Z3, and a post-workshop survey ($N=21$) shows that students deemed \tool easy to use and helpful for learning.
Our evaluation is limited in its formative nature;
however, the lessons we learned from our design process, the workshop, and our year-long experience using the tool imply several directions for improving \tool and further evaluations.
We hope that our findings can inform future educational tools and their design guidelines for logic modeling and programming education.

\begin{acks}
We thank all our study participants for their invaluable input.
% We also thank our colleagues and the reviewers for their feedback.
% Acknowledgements may include funding information and are omitted for the double-blind review process.
\end{acks}
\end{sloppypar}


\bibliographystyle{ACM-Reference-Format}

\bibliography{citations}

\end{document}